\documentclass[journal]{IEEEtran}

\usepackage{bm,bbm}
\usepackage{amsfonts,amsmath}
\usepackage{amssymb}
\usepackage{algorithm,algorithmicx}
\usepackage{algpseudocode}
\usepackage{graphicx,subfigure}

\ifCLASSINFOpdf
\else
\fi
\newtheorem{assumption}{Assumption}
\newtheorem{lemma}{Lemma}
\newtheorem{theorem}{Theorem}

\newtheorem{remark}{Remark}
\newtheorem{definition}{Definition}
\newtheorem{corollary}{Corollary}

\begin{document}
%
\title{Strong Consistency of Spectral Clustering for the Sparse Degree-Corrected Hypergraph Stochastic Block Model}
%
%
%

\author{Chong~Deng,~Xin-Jian~Xu,~and Shihui~Ying,~\IEEEmembership{Member,~IEEE}
						
	\thanks{C. Deng is with the Department of Mathematics, Shanghai University, Shanghai 200444, China (e-mail: dengchong@shu.edu.cn).}
	\thanks{X.-J. Xu is with Qianweichang College, Shanghai University, Shanghai 200444, China (e-mail: xinjxu@shu.edu.cn).}
	\thanks{S. Ying is with the Department of Mathematics, Shanghai University, Shanghai 200444, China (e-mail: shying@shu.edu.cn).}
						
}

%
%

\markboth{Journal of \LaTeX\ Class Files}%
{Shell \MakeLowercase{\textit{et al.}}: Bare Demo of IEEEtran.cls for IEEE Journals}
%



\maketitle

\begin{abstract}
	We prove strong consistency of spectral clustering under the degree-corrected hypergraph stochastic block model in the sparse regime where the maximum expected hyperdegree is as small as $\Omega(\log n)$ with $n$ denoting the number of nodes. We show that the basic spectral clustering without preprocessing or postprocessing is strongly consistent in an even wider range of the model parameters, in contrast to previous studies that either trim high-degree nodes or perform local refinement. At the heart of our analysis is the entry-wise eigenvector perturbation bound derived by the \lq\lq leave-one-out\rq\rq~technique. To the best of our knowledge, this is the first entry-wise error bound for degree-corrected hypergraph models, resulting in the strong consistency for clustering non-uniform hypergraphs with heterogeneous hyperdegrees.
\end{abstract}


%
\IEEEpeerreviewmaketitle

\section{Introduction}

\IEEEPARstart{C}{ommunity} detection, also known as graph clustering, occupies a central position in modern network science. The goal is to partition nodes in a network into communities such that the nodes in the same community are more similar than those belonging to different communities. Among numerous algorithms designed for this purpose, the model-based algorithms stand out for their statistical guarantees. The stochastic block model (SBM)~\cite{holland1983} is one of the well known generative models for random graphs with community structure. In this model, edges are independently generated with probabilities depending only on the community memberships of nodes, and thus the nodes in the same community have the same expected degrees. As a consequence, the SBM delivers a poor description of real networks that exhibit degree heterogeneity even within communities. To overcome this limitation, the degree-corrected stochastic block model (DCSBM)~\cite{karrer2011} has been proposed by introducing a set of node-specific parameters to allow any possible degree distribution. It shows that the consideration of the degree heterogeneity significantly improves the ability of the model to fit real networks.

For community detection in a generative model, i.e., recovering the hidden community assignments from a single instance generated by the model, it is usually concerned with consistency of an algorithm:
\begin{itemize}
	\item \textit{strong consistency} (or exact recovery). Finding the true partition of all nodes (up to a permutation) with high probability;
	      	      	      	      	      	
	\item \textit{weak consistency} (or almost exact recovery). Finding the true partition of all but a vanishing fraction of nodes (up to a permutation) with high probability.
\end{itemize}
The fundamental limits of community detection under the (DC)SBM have been widely studied~\cite{zhao2012,krzakala2013,gao2018,gulikers2018}, and the performance of many clustering algorithms, such as spectral algorithms~\cite{rohe2011,lei2015}, has also been investigated both theoretically and numerically. However, as graphs model pairwise relationships by edges connecting pairs of nodes, they fail to capture higher-order interactions, and this is where hypergraphs come into play. Different from graphs, edges in a hypergraph, called hyperedges, can connect any number of nodes. Due to this merit, community detection in hypergraphs, or hypergraph clustering, has received increasing attention recently. In particular, the SBM has been generalized to the hypergraph SBM (HSBM)~\cite{chodrow2021}, in which the probability of occurrence of a hyperedge depends on the community memberships of all nodes in it. Considering the degree heterogeneity, the DCSBM has also been generalized to the degree-corrected HSBM (DCHSBM). Yet, there are very few studies on the theoretical limits for community detection under the DCHSBM and the algorithms achieving the detectability thresholds.

As one of the most popular algorithms, spectral clustering has been adopted for graph clustering. In its basic form, nodes are first represented as points in a low-dimensional space based on the leading eigenvectors, and these points are then clustered using standard clustering algorithms such as $k$-means. To develop spectral algorithms for hypergraph clustering, several matrices have been considered including adjacency matrix~\cite{ahn2018}, hypergraph Laplacian~\cite{ghoshdastidar2017} and adjacency tensor~\cite{ke2019}. All these algorithms are weakly consistent in the HSBM under certain conditions on the model parameters. As for strong consistency, Cole and Zhu~\cite{cole2020} achieved exact recovery for dense hypergraphs but it is suboptimal in the sparse case. On the other hand, two-stage algorithms have been proposed~\cite{ahn2018,chien2019,zhang2021}: first obtain a weakly consistent community assignment from the trimmed adjacency matrix or hypergraph Laplacian and then perform local refinement to guarantee strong consistency.

For graph clustering, it has been proved that the basic spectral clustering without trimming or refinement is strongly consistent~\cite{su2019,abbe2020}. Therefore, it is natural to ask a question: whether the basic hypergraph spectral clustering could achieve strong consistency or not? Answering this question requires a more refined entry-wise analysis of eigenvector perturbations. The goal of this paper is to prove strong consistency of the basic spectral clustering under the DCHSBM in the sparse regime where the maximum expected hyperdegree might be of order $\Omega(\log n)$.

\subsection{Main contributions}
Let $A$ be the weighted adjacency matrix (Equation \eqref{wadj}) of a hypergraph generated by a DCHSBM (Definition \ref{defdchsbm}) and $P=\mathbb{E}[A]$ be the population counterpart of $A$. Let $\hat{U}$ and $U$ be the matrices respectively formed by stacking the $K$ leading eigenvectors of $A$ and $P$, where $K$ is the number of communities.
As the matrix $U$ contains all information about the true community assignment, the task becomes bounding the deviation between $\hat{U}$ and $U$. While the well known Davis-Kahan theorem specifies an upper bound on $\lVert \hat{U}\hat{O}-U\rVert_F$ for some orthogonal matrix $\hat{O}$ (this matrix is used to align $\hat{U}$ and $U$; see Equation \eqref{sgnmat} for its specific expression), it gives a loose bound on $\lVert \hat{U}\hat{O}-U\rVert_{2,\infty}$, hence no guarantee for node-wise behavior. We show that simply applying $k$-means on the row-normalized leading eigenvector matrix of the weighted adjacency matrix achieves strong consistency in an even wider range of parameters. We develop a sharp deviation bound on $\lVert A-P\rVert$ for non-uniform hypergraphs with general edge probabilities. As a byproduct, we also derive an upper bound on mis-clustered nodes incurred by an approximate $k$-means algorithm which leads to weak consistency of Algorithm~\ref{alg1} (see Section~\ref{algorithm}). Via the leave-one-out analysis, we obtain an eigenvector perturbation bound in two-to-infinity norm which is the first result on node-wise error bounds for hypergraph models. With this bound, we also study strong consistency of Algorithm~\ref{alg2} (see Section~\ref{algorithm}), which is simple in nature but will be extremely hard to analyze if one only has a bound on the deviation between $\hat{U}$ and $U$ in the Frobenius norm. Finally, we consider a special case of the model where the conditions for exact recovery will be expressed more clearly. To the best of our knowledge, our study gives the first strong consistency result for clustering non-uniform hypergraphs with heterogeneous hyperdegrees.

\subsection{Related Work}
To date, there are few studies on strong consistency for hypergraph stochastic block models. Kim et al~\cite{kim2018} demonstrated that exact recovery shows a sharp phase-transition behavior for the uniform HSBM with two equal-sized and symmetric communities. They proposed a semidefinite programming algorithm which is strongly consistent in an order-wise optimal parameter regime. Ahn et al~\cite{ahn2018} investigated consistency of spectral clustering in weighted uniform HSBMs where the number of clusters $K$ is constant. Cole and Zhu~\cite{cole2020} proposed a spectral algorithm based on the hypergraph adjacency matrix and proves that the algorithm achieves exact recovery in the dense uniform HSBM where $K=\Theta(\sqrt{n})$. Chien et al~\cite{chien2019} showed that spectral algorithms with local refinement achieve the exact recovery criteria in the sparse HSBM. Zhang and Tan~\cite{zhang2021} studied fundamental limits of exact recovery in the general uniform HSBM and develop a two-stage algorithm that meets the achievability threshold.

In \cite{abbe2020}, the authors performed an entry-wise eigenvector analysis and proved that spectral clustering achieves the threshold of exact recovery in a graph SBM with two blocks. In our study, we adopt the idea of the leave-one-out technique. Compared with \cite{abbe2020}, our main contributions are as follows. First, we generalize the entry-wise eigenvector perturbation analysis for graphs to hypergraphs. This is non-trivial because the weighted adjacency matrix $A$ violates the \lq\lq row- and column-wise independence\rq\rq~assumption made in \cite{abbe2020}. Based on the idea of the leave-one-out method, we introduce a set of suitably defined surrogate matrices $A^{(l)}$ where randomness contributed by hyperedges containing $l$ is eliminated, and then resort to the matrix Bernstein inequality where sequential transformations and inequalities are performed to address the dependency across entries of $A$. Second, we develop a sharp deviation bound on $\lVert A-P\rVert$ for non-uniform hypergraphs. Finally, we derive the strong consistency result for spectral clustering in the non-uniform DCHSBM with multiple communities.

Gaudio and Joshi~\cite{Gaudio2022} derived an entry-wise bound on the second eigenvector of the adjacency matrix and prove strong consistency of spectral clustering for the uniform HSBM with two communities. Compared with~\cite{Gaudio2022}, we consider different algorithms and have a different goal: we try to find under what conditions on degree heterogeneity, the number of communities, sparsity and the minimal non-zero eigenvalue of the matrix (See Lemma~\ref{lma1}), spectral clustering without pre-processing and post-processing could achieve strong consistency. Gaudio and Joshi~\cite{Gaudio2022} instead aimed at proving their algorithm can achieve the theoretical threshold. While our settings are more general, our result does not achieve the theoretical limit in this special case. Both our work and Ref.~\cite{Gaudio2022} perform the entry-wise eigenvector analysis. The main difference is as follows. Ref.~\cite{Gaudio2022} obtains a more refined $l_{\infty}$ eigenvector perturbation bound by generalizing Theorem 1.1 in Ref.~\cite{abbe2020} to the two-block HSBM case. We derive an entry-wise bound for deviation between $\hat{U}$ and $U$ in $l_{2,\infty}$ norm by applying the matrix Bernstein inequality together with the Davis-Kahan theorem and the triangle inequality, where we carefully handle the dependence across the entries of $A$. Our entry-wise bound makes it possible to guarantee strong consistency of the considered algorithms in the general $K$-block case, where $K$ is even allowed to diverge to infinity at a slow rate.

Wang~\cite{Wang2023} established the information-theoretical threshold for strong consistency in non-uniform HSBM with two equal-sized communities.
However, the algorithms considered in Ref.~\cite{Wang2023} are not applicable to the case of multiple blocks.
Dumitriu and Wang~\cite{Dumitriu2023} derived sharp threshold for exact recovery in non-uniform HSBM with multiple communities and provided multi-stage algorithms that successfully achieve exact recovery above the threshold.
Compared with Ref.~\cite{Dumitriu2023}, we consider strong consistency of spectral clustering without preprocessing or
postprocessing and derive the first entry-wise eigenvector perturbation bound for non-uniform hypergraphs with heterogeneous hyperdegrees.

\subsection{Paper Organization}
We start in Section~\ref{preliminaries} with an introduction of the DCHSBM, the weighted adjacency matrix and two hypergraph clustering algorithms. We describe our main results as well as its proof in Section~\ref{mainresult}. The consistency results in a special case is given in Section~\ref{specialcases}. Section~\ref{conclusion} contains concluding remarks.

\subsection{Notations}
Given a matrix $X$, we use $X_{i\cdot}$ to refer the $i$-th row of $X$. Let $\lVert \cdot\rVert$ denote the spectral norm of a matrix or the $l_2$ norm of a vector and $\lVert\cdot\rVert_{F}$ denote the matrix Frobenius norm. Let $\lVert X\rVert_{2,\infty}=\max_{i}\lVert X_{i\cdot}\rVert$ be the two-to-infinity norm of $X$. For any vector $v\in\mathbb{R}^n$, let $v_{max}=\max_{i}v_i$, $v_{min}=\min_{i}v_i$ and $\mathrm{diag}(v)$ be an $n\times n$ matrix with zero off-diagonal whose $ii$-th element is $v_i$. For any positive integer $n$, we use $[n]=\{1,2,\cdots,n\}$ to denote the set of positive integers not greater than $n$. Denote $\delta$ as the Kronecker function: for $i,j\in\mathbb{Z}$, $\delta_{i,j}=1$ if $i=j$, and $\delta_{ij}=0$ otherwise. Throughout the paper, we use the standard asymptotic notations: $o(\cdot),O(\cdot),\Theta(\cdot),\Omega(\cdot)$ and $\omega(\cdot)$.

\section{Preliminaries}\label{preliminaries}

\subsection{The Model}\label{model}

Let $n$ be the number of nodes in a hypergraph. The nodes are divided into $K$ communities and each node $i$ is assigned a label $g_i\in[K]$ representing the community to which it belongs. Additionally, each node $i$ is assigned a parameter $\theta_i>0$ to control its expected degree. Let $M\geqslant2$ be the maximum hyperedge cardinality in the hypergraph. Let $E$ be the hyperedge set of a complete hypergraph on $[n]$ with edge size between 2 and $M$. Each possible hyperedge $e\in E$ (hyperedge that contains duplicate nodes is allowed) is associated with an indicator variable $h_e$ such that $h_e=1$ if $e$ is present and $h_e=0$ otherwise.

\begin{definition}[Degree-corrected hypergraph stochastic block model]\label{defdchsbm}
	In a DCHSBM, $\{h_e\}_{e\in E}$ are independent Bernoulli random variables satisfying
	\begin{equation*}
		\mathbb{P}(h_e=1)=b_e\pi(\theta_e)\Phi(g_e)\in[0,1],
	\end{equation*}
	where $\pi(\theta_e)=\prod_{i\in e}\theta_i$ is the product of the hyperdegree parameters of the nodes in $e$, $\Phi$ is an affinity function that maps the group assignment to a non-negative real number and the coefficient $b_e$ denotes the number of distinct ways to order the nodes of $e$.
\end{definition}

For example, suppose $n=8, K=2, M=5$ and $g=(1,1,1,1,1,2,2,2)$. For three possible hyperedges $e_1=(2,6), e_2=(1,3,8), e_3=(1,4,4,7)$, we have $b_{e_1}=2,b_{e_2}=6,b_{e_3}=12$ and thus $\mathbb{P}(h_{e_1}=1)=2\theta_2\theta_6\Phi(1,2)$, $\mathbb{P}(h_{e_2}=1)=6\theta_1\theta_3\theta_8\Phi(1,1,2)$ and $\mathbb{P}(h_{e_3}=1)=12\theta_1\theta_4^2\theta_7\Phi(1,1,1,2)$.

Let $n_k$ denote the size of the $k$-th community. Without loss of generality, we assume $n_1\geqslant n_2\geqslant\cdots\geqslant n_K$. The community memberships can also be represented by an assignment matrix $Z\in\{0,1\}^{n\times K}$ where $Z_{ik}=1$ if $g_i=k$ and $Z_{ik}=0$ otherwise. For each $e\in E$, we define $g_e$ as a vector of the cluster labels for nodes in $e$.

To ensure model identifiability, we impose the constraint that $\sum_{i}\theta_i\delta_{g_i,k}=n_k$ for each $k\in[K]$. Thus, the DCHSBM nests the HSBM as a special case by setting $\theta_i=1$ for all $i\in[n]$. Similarly, we define $\theta_e$ as a vector of the hyperdegree parameters of the nodes in $e$.

Furthermore, we introduce vectors $a_e\in\mathbb{R}^n$, where $a_{ei}$ is the number of occurrences of node $i$ in $e$. The hyperdegree of node $i$ is defined by $d_i=\sum_{e\in E}a_{ei}h_e$ and the cardinality of each possible hyperedge $e\in E$ is given by $|e|=\sum_{i=1}^na_{ei}$.

A basic but important choice of $\Phi$ is the so-called \lq\lq all-or-nothing\rq\rq~affinity function, where the hyperedge probability depends on whether all nodes in the hyperedge are in the same community or not.

\begin{definition}[Degree-corrected hypergraph planted partition model]\label{defdchppm}
	For each $e=\{i_1,\cdots,i_m\}\in E$, it is generated with probability
	\begin{equation*}
		\mathbb{P}(h_e=1)=\alpha_mb_e\pi(\theta_e)\cdot((p-q)\mathbbm{1}_{\{g_{i_1}=\cdots=g_{i_m}\}}+q),
	\end{equation*}
	where $p>q>0$ are constants independent of $n$ and $\alpha_m\geqslant0$ is a scaling factor that varies with $n$ and controls the number of the hyperedge cardinality $m$.
\end{definition}

We investigate consistency of spectral algorithms for this model in Section~\ref{specialcases} as a case study and test the performance of the considered algorithms in this model in Section~\ref{experiment}.

\subsection{The Weight Adjacency Matrix}\label{adjmatrix}
Spectral algorithms for community detection in the model depend on the weighted adjacency matrix $A\in\mathbb{R}^{n\times n}$ which is non-negative symmetric with entries:
\begin{equation}\label{wadj}
	A_{ij}=
	\begin{cases}
		\sum_{e\in E}\frac{a_{ei}a_{ej}}{|e|-1}h_e,     & \text{ if }i\neq j, \\
		\sum_{e\in E}\frac{a_{ei}(a_{ei}-1)}{|e|-1}h_e, & \text{ if }i=j.
	\end{cases}
\end{equation}
When there is no hyperedge containing repeated nodes, the diagonal entries of $A$ are all zero. The so-defined weighted adjacency matrix corresponds to the adjacency matrix of a weighted undirected graph projected by a hypergraph~\cite{carletti2021}. In this work, we investigate consistency of the hypergraph spectral clustering based on the weighted adjacency matrix in the DCHSBM.

To understand why the spectrum of $A$ contains information about the hidden community structure, we first take a close look at its expectation $P=\mathbb{E}[A]$. Denote the eigenvalues of $P$ by $\{\lambda_i\}_{i=1}^n$ such that $|\lambda_1|\geqslant\cdots\geqslant|\lambda_n|$ and let $\{u_i\}_{i=1}^n$ be the corresponding unit-norm eigenvectors. Define $U=(u_1,\cdots,u_K)\in\mathbb{R}^{n\times K}$ and $U^*$ the normalized matrix with $U_{i\cdot}^*=U_{i\cdot}/\lVert U_{i\cdot}\rVert$. The following lemma characterizes the eigen-structure of $P$.

\begin{lemma}[The eigen-structure of $P$]\label{lma1}
	~
	\begin{itemize}
		\item[(i).]
		      There exists a symmetric matrix $B\in\mathbb{R}^{K\times K}$ such that
		      \begin{equation}\label{lma1eq1}
		      	P=\mathrm{diag}(\theta) ZBZ^T\mathrm{diag}(\theta).
		      \end{equation}
		      		      		      		      		      		
		\item[(ii).]
		      If $B$ is full rank, then there exists an orthogonal matrix $Q\in\mathbb{R}^{K\times K}$ such that for any $i\in[n]$,
		      \begin{equation}\label{lma1eq2}
		      	U_{i\cdot}={\tilde{\theta}_i}Q_{g_i\cdot},
		      \end{equation}
		      where $\tilde{\theta}_i=\theta_i/\phi_{g_i}$ and $\phi_k=\sqrt{\sum_{j=1}^n\theta_j^2\delta_{g_j,k}}$ for $k\in[K]$.
	\end{itemize}
\end{lemma}

As an immediate consequence of~\eqref{lma1eq1}, we have $\mathrm{rank}(P)\leqslant K$, which means $\lambda_{K+1}=\cdots=\lambda_n=0$. Readers familiar with spectral clustering in graphs will find that $P$ could be regarded as the population adjacency matrix of an ordinary DCSBM parameterized by $g$, $\theta$ and $B$, which could be viewed as a \lq\lq projected model\rq\rq. It is known that graph projection may cause the community structure unidentifiable under some parameter space~\cite{ghoshdastidar2017,zhang2021}. Therefore, the condition that matrix $B$ is full rank, i.e., $|\lambda_K|>0$, is crucial for the success of the graph-projection-based clustering algorithms.

According to~\eqref{lma1eq2}, for any two nodes $i$ and $j$, if they belong to the same community, i.e., $g_i=g_j$, then $U_{i\cdot}$ and $U_{j\cdot}$ point to the same direction in $\mathbb{R}^K$; otherwise, $U_{i\cdot}$ and $U_{j\cdot}$ are orthogonal to each other. By normalizing the rows of $U$ to have unit length, we have
\begin{equation*}
	\lVert U_{i\cdot}^*-U_{j\cdot}^*\rVert=\sqrt{2}\cdot\mathbbm{1}_{\{g_i\neq g_j\}}.
\end{equation*}
Thus, the community memberships of all nodes will be exactly recovered from $U^*$ without difficulty.

In practice, $P$ is not observed and spectral clustering is applied to the noisy observation $A$. Thus, if $\hat{U}^*$ is close enough to $U^*$, we can expect that the spectral algorithms could still successfully recover the true community memberships. However, bounding the deviation of $\hat{U}^*$ from $U^*$ (especially entry-wise) turns out to be a non-trivial task once the entries of $A$ are no longer independent. More care must be taken when dealing with the complex dependency across entries.

\subsection{Spectral Algorithms}\label{algorithm}
Denote the $K$ leading eigenvalues of $A$ by $\{\hat{\lambda}_k\}_{k=1}^K$ such that $|\hat{\lambda}_1|\geqslant\cdots\geqslant|\hat{\lambda}_K|$ and let $\{\hat{u}_k\}_{k=1}^K$ be the corresponding unit-norm eigenvectors. Let $\hat{U}\in\mathbb{R}^{n\times K}$ be the matrix that contains the $K$ leading eigenvectors of $A$ as columns and $\hat{U}^*\in\mathbb{R}^{n\times K}$ be the row-normalized version of $\hat{U}$, i.e., $\hat{U}^*_{i\cdot}=\hat{U}_{i\cdot}/\lVert \hat{U}_{i\cdot}\rVert$ for all $i\in[n]$.\par

The first algorithm we consider is the classical spectral clustering summarized in Algorithm~\ref{alg1}. For strong consistency of the $k$-means step, we utilize the result of Theorem 2.3 in Ref.~\cite{su2019}. The second one is a simple thresholding algorithm listed in Algorithm~\ref{alg2}. Two nodes will be assigned to the same community when the distance between the corresponding rows of $\hat{U}^*$ is small enough. Concretely, we try to find a threshold $\tau$ such that, starting from an empty graph $G=([n],\emptyset)$, after connecting all node pairs $(i,j)$ satisfying $\lVert\hat{U}_{i\cdot}^*-\hat{U}_{j\cdot}^*\rVert< \tau$, $G$ has exactly $K$ connected components, which corresponds to the true $K$ communities. Initially there are $n$ connected components, each of which contains only one node. After adding an edge, the number of components either decreases by one or remains the same. If we connect all node pairs, there will be only one connected component. Therefore, Algorithm \ref{alg2} can successfully output a partition of $[n]$ into $K$ communities and the output is unique.

\begin{algorithm}[H]
	\caption{Hypergraph spectral clustering with $k$-means}\label{alg1}
	\begin{algorithmic}[1]
		\Require The hypergraph $\mathcal{H}$ and the number of communities $K$.
		\State Construct the weighted adjacency matrix $A$ and compute $\hat{U}^*$.
		\State Run $k$-means algorithm with $k=K$ on the rows of $\hat{U}^*$.
		\Ensure A community assignment $\hat{g}$ where $\hat{g}_i$ is the cluster index of $i$-th row.
	\end{algorithmic}
\end{algorithm}

\begin{algorithm}[H]
	\caption{Hypergraph spectral clustering via thresholding}\label{alg2}
	\begin{algorithmic}[1]
		\Require The hypergraph $\mathcal{H}$ and the number of communities $K$.
		\State Construct the weighted adjacency matrix $A$ and compute $\hat{U}^*$.
		\State Let $G$ be a simple graph with node set $[n]$ and an empty edge set.
		\State Sort node pairs $\{(i,j)|i,j\in[n]\}$ in ascending order by $\lVert\hat{U}_{i\cdot}^*-\hat{U}_{j\cdot}^*\rVert$.
		\State Add edges to node pairs in $G$ in the above order until $G$ has exactly $K$ connected components.
		\Ensure A community assignment $\tilde{g}$ where $\tilde{g}_i$ is the index of the component to which $i$ belongs in $G$.
	\end{algorithmic}
\end{algorithm}

For both algorithms, the successful recovery of community labels of all nodes depends strongly on $\hat{U}^*$ and $U^*$ being sufficiently close entry-wise. When proving strong consistency of Algorithm \ref{alg2}, we show that with high probability: if $\lvert\hat{U}_{i\cdot}^*-\hat{U}_{j\cdot}^*\rvert<1/\sqrt{2}$ then $g_i=g_j$, otherwise $g_i\neq g_j$. Here we comment on the threshold $1/\sqrt{2}$, which could be replaced by any constant $\tau\in(0,\sqrt{2})$. For any nodes $i$ and $j$, by the triangle inequality, we have
$\lVert\hat{U}_{i\cdot}^*-\hat{U}_{j\cdot}^*\rVert\leqslant2\lVert\hat{U}^*\hat{O}-U^*\rVert_{2,\infty}$ if $g_i=g_j$ and
$\lVert\hat{U}_{i\cdot}^*-\hat{U}_{j\cdot}^*\rVert\geqslant\sqrt{2}-2\lVert\hat{U}^*\hat{O}-U^*\rVert_{2,\infty}$ otherwise. Since both bounds are tight in the worst case, a sufficient and necessary condition to ensure strong consistency of Algorithm~\ref{alg2} is
\begin{equation}\label{eq231}
	\lVert\hat{U}^*\hat{O}-U^*\rVert_{2,\infty}<\frac{1}{2}\min\{\tau,\sqrt{2}-\tau\}.
\end{equation}
Because $1/\sqrt{2}$ maximizes the right-hand side of~\eqref{eq231}, it is optimal in the sense that it imposes the mildest requirement on $\lVert\hat{U}^*\hat{O}-U^*\rVert_{2,\infty}$.

\section{Consistency of hypergraph spectral clustering}\label{strongconsistency}

\subsection{The Main Result}\label{mainresult}
We show that under mild conditions on the model parameters, both algorithms mentioned above are strongly consistent.
Let $g'$ be an estimator of $g$ that partitions nodes into $K$ communities. The number of mis-clustered nodes of $g'$ is defined by
\begin{equation*}
	l(g,g')=\min_{\sigma\in\mathcal{S}_K}\sum_{i=1}^n\mathbbm{1}_{\{g_i\neq\sigma(g_i')\}},
\end{equation*}
where $\mathcal{S}_K$ is the $K$-th order symmetric group and the minimum is taken over all possible permutation of $[K]$. We say $g'$ is weakly consistent if $l(g,g')=o(n)$ with probability $1-o(1)$, which only needs the fraction of mis-clustered nodes to vanish in the large $n$ limit. In contrast, strong consistency requires $l(g,g')=0$ with probability $1-o(1)$, which means that the community membership of all nodes should be exactly identified in large samples. For notational simplicity, we write \lq\lq $g'=g$ with high probability\rq\rq~in short.

To ensure strong consistency of spectral algorithms, we make the following assumptions about the model parameters.

\begin{assumption}\label{asp1}
	$M=O(1)$.
\end{assumption}

That is, $M$ is not allowed to vary with $n$. A constant maximum hyperedge cardinality is necessary for both the sharp upper bound on $\lVert A-P\rVert$ and the bound on the node-wise deviation $\lVert \hat{U}\hat{O}-U\rVert_{2,\infty}$ to hold. This is a mild condition for $M$ since in practice $M$ is usually much smaller than $n$.

\begin{assumption}\label{asp2}
	$n_1/n_K=O(1)$.
\end{assumption}

In other words, the nodes form communities of fairly balanced size, which is a standard assumption for strong consistency~\cite{ahn2018,zhang2021,su2019}.

\begin{assumption}\label{asp3}
	$\kappa={|\lambda_1|}/{|\lambda_K|}=O(\sqrt{\log n})$.
\end{assumption}

This is a necessary condition for Theorem~\ref{thm3} in Section~\ref{proof}. In the literature, $\kappa$ is usually assumed to be bounded from above by a constant~\cite{abbe2017}. A weaker requirement is made here thanks to the sharp deviation bound on $\lVert A-P\rVert$.

\begin{assumption}\label{asp4}
	$B$ is full rank, or equivalently, $\lambda_K\neq0$.
\end{assumption}

This assumption is in fact a critical condition for the identifiability of the partitions. Since the hypergraph is projected to a weighted graph, a large eigen-gap, i.e., a large value of $|\lambda_K|$, ensures that the partition is identifiable and thus could be successfully extracted. It is unclear whether Algorithms 1 and 2 are still strongly consistent under milder conditions on model parameters (e.g., $B$ contains distinct rows).

Let $d= \max\{n\max_{ij}P_{ij},c_0\log n\}$ be an upper bound of the expected node hyperdegrees for some constant $c_0>0$. Define
\begin{equation}
	\gamma=\frac{\max_{i\in[n]}\lVert U_{i\cdot}\rVert}{\min_{i\in[n]}\lVert U_{i\cdot}\rVert}=\frac{\tilde{\theta}_{max}}{\tilde{\theta}_{min}}.
\end{equation}
$\gamma$ depends only on the model parameter $\theta$ and specifies an upper bound on $\lVert U_{i\cdot}\rVert/\lVert U_{j\cdot}\rVert$, which is useful when characterizing the upper bound for clustering error rate. For a HSBM, we have $\gamma=\sqrt{n_1/n_K}=O(1)$ due to Assumption 2, which corresponds to the absence of degree heterogeneity in the HSBM. In general, the stronger the degree heterogeneity is, the larger $\gamma$ is and the harder for clustering algorithms to achieve exact recovery.

The main result is given by the following theorem.

\begin{theorem}\label{thm1}
	Let $A$ be a weighted adjacency matrix of a hypergraph generated by a DCHSBM that Assumptions~\ref{asp1}-\ref{asp4} hold.
	\begin{itemize}
		\item[(i).]
		      There exists a constant $C_1=C_1(M,c_0)$ such that if $n$ is sufficiently large\footnote{Note that the number of nodes $n$ being sufficiently large is a necessary condition for the success of the $k$-means algorithm~\cite{ghoshdastidar2017,su2019}.} and
		      \begin{equation}\label{condition1}
		      	\frac{\gamma K^{3/2}\sqrt{d\log n}}{|\lambda_K|}\leqslant C_1,
		      \end{equation}
		      then $\hat{g}=g$ with high probability.
		      		      		      		      		      		
		\item[(ii).]
		      There exists a constant $C_2=C_2(M,c_0)$ such that if
		      \begin{equation}\label{condition2}
		      	\frac{\gamma \sqrt{d\log n}}{|\lambda_K|}\leqslant C_2,
		      \end{equation}
		      then $\tilde{g}=g$ with high probability.
	\end{itemize}
\end{theorem}

Though the condition~\eqref{condition2} is milder than~\eqref{condition1}, Algorithm~\ref{alg1} may have better theoretical and/or practical performance than Algorithm~\ref{alg2} since Theorem~\ref{thm1} only provides sufficient conditions for the considered algorithms to achieve exact recovery. Note that conditions~\eqref{condition1} and~\eqref{condition2} do not directly reveal the fact that strong consistency should be more achievable when a hypergraph gets denser since both $d$ and $\lambda_K$ are affected by the sparsity of the hypergraph. To better demonstrate this, we consider the consistency of $\hat{g}$ and $\tilde{g}$ in the $m$-uniform HPPM, which has been widely studied in the literature~\cite{ahn2018,cole2020,kim2018}. In the $m$-uniform HPPM, all hyperedges are of size $m$ and the generation probability of $e=\{i_1,\cdots,i_m\}$ is $\mathbb{P}(e)=\alpha_mb_e((p-q)\cdot\mathbbm{1}_{\{g_{i_1}=\cdots=g_{i_m}\}}+q)$.

\begin{corollary}\label{coro1}
	Let $A$ be a weighted adjacency matrix of a hypergraph generated by the $m$-uniform HPPM, then Assumption \ref{asp4} holds. If
	\begin{equation*}
		K=O((\log n)^{\frac{1}{2m-2}}),
	\end{equation*}
	then Assumption \ref{asp3} holds. When Assumptions \ref{asp1}-\ref{asp4} all hold, we have the following conclusions:
						
	\begin{itemize}
		\item[(i).]
		      There exists a constant $C_1>0$ such that for sufficiently large $n$, if
		      \begin{equation*}
		      	\alpha_m\geqslant C_1\frac{K^{2m+1}\log n}{n^{m-1}},
		      \end{equation*}
		      then $\hat{g}=g$ with high probability.
		      		      		      		      		      		
		\item[(ii).]
		      There exists a constant $C_2>0$ such that if
		      \begin{equation*}
		      	\alpha_m\geqslant C_2\frac{K^{2m-2}\log n}{n^{m-1}},
		      \end{equation*}
		      then $\tilde{g}=g$ with high probability.
	\end{itemize}
\end{corollary}

This corollary is a special case of Corollary~\ref{coro2} in Section~\ref{uniform}. The proof could be found in Appendix~\ref{proofcoro2}. We compare Algorithms~\ref{alg1} and~\ref{alg2} with existing strongly consistent algorithms in Table~\ref{tab1}. While~\cite{ahn2018,chien2019,zhang2021} require the number of communities $K$ to be a constant, \cite{cole2020} allows $K$ to grow like $O(\sqrt{n})$ which is much faster than ours, but that algorithm only works for dense hypergraphs. In contrast, we allow $K$ to diverge to infinity at a slow rate in the sparse regime. When $K$ is a constant, the sparsity required by Algorithms~\ref{alg1} and~\ref{alg2} meets the theoretical limit $\Omega(\log n/n^{m-1})$.

\begin{table*}
	\caption{Comparison of strongly consistent algorithms}
	\label{tab1}
	\begin{center}
		\begin{tabular}{c|ccccc}
			\hline \hline
			Paper                & Algorithm type & $m$    & $K$                            & $\alpha_m$                               & Sizes of blocks \\
			\hline
			\cite{kim2018}       & SDP            & $O(1)$ & 2                              & $\Omega(\log n/{n-1\choose m-1})$        & Equal           \\
			\cite{ahn2018}       & Spectral+LR    & $O(1)$ & $O(1)$                         & $\Omega(n\log n/{n\choose m})$           & Almost equal    \\
			\cite{chien2019}     & Spectral+LR    & $O(1)$ & $O(1)$                         & $\Omega(\log n/n^{m-1})$                 & Almost equal    \\
			\cite{cole2020}      & Spectral       & $O(1)$ & $O(\sqrt{n})$                  & $\Theta(1)$                              & Equal           \\
			\cite{zhang2021}     & Spectral+LR    & $O(1)$ & $O(1)$                         & $\Theta({\log n}/{n^{m-1}})$             & Almost equal    \\
			Algorithm~\ref{alg1} & Spectral       & $O(1)$ & $O((\log n)^{\frac{1}{2m-2}})$ & $\Omega(\frac{K^{2m+1}\log n}{n^{m-1}})$ & Almost equal    \\
			Algorithm~\ref{alg2} & Spectral       & $O(1)$ & $O((\log n)^{\frac{1}{2m-2}})$ & $\Omega(\frac{K^{2m-2}\log n}{n^{m-1}})$ & Almost equal    \\
			\hline
		\end{tabular}
	\end{center}
\end{table*}

\subsection{Proof of the main result}\label{proof}

The outline of the proof is as follows. We first derive a sharp bound on the deviation $\lVert A-P\rVert$ by the combinatorial technique~\cite{lei2015}. Then, we obtain an upper bound on $\lVert \hat{U}\hat{O}-U\rVert_{2,\infty}$ using the leave-one-out technique~\cite{abbe2020}. Finally, we bound $\lVert \hat{U}^*\hat{O}-U^*\rVert_{2,\infty}$ and analyze the performance of Algorithms~\ref{alg1} and~\ref{alg2}.

To derive a tight bound on $\lVert A-P\rVert$, we adopt the Kahn-Szemeredi argument~\cite{Friedman1989}, which has been used to bound the second largest eigenvalue of ER graphs~\cite{feige2005} and the spectral norm of general binary symmetric random matrices~\cite{lei2015}.

\begin{theorem}\label{thm2}
	Let Assumption \ref{asp1} hold. For any $r>0$, there exists a constant $C=C(M,c_0,r)$ such that
	\begin{equation*}\lVert A-P\Vert\leqslant C\sqrt{d}\end{equation*}
		with probability at least $1-O(n^{-r})$.
		\end{theorem}
												
		It should be mentioned that~\cite{ahn2018} first extended the above techniques to hypergraphs and obtained a concentration bound for the weighted uniform hypergraph stochastic block model in which the edge weight has binary expected value. On the contrary, the model we consider here is non-uniform and the edge probabilities are more general. In this scenario, the DCHSBM allows variation of density of edges of different cardinality. Since we focus on sparse hypergraphs, the hyperdegrees of nodes, which could be as small as $\Omega(\log n)$, will be regarded as a summary of the sparsity of edges of different cardinalities.
												
		As a byproduct, we establish the weak consistency of Algorithm~\ref{alg1} based on Theorem~\ref{thm2}. The classical way to derive the error bound of Algorithm~\ref{alg1} consists of three steps: (i) Bound $\lVert A-P\rVert$; (ii) Bound $\lVert \hat{U}\hat{Q}-U\rVert_F$ and $\lVert \hat{U}^*\hat{Q}-U^*\rVert_F$ for some orthogonal matrix $\hat{Q}\in\mathbb{R}^{K\times K}$; and (iii) Bound the error incurred by an $(1+\epsilon)$-approximate $k$-means algorithm~\cite{kumar2004simple}. Combining Theorem~\ref{thm2}, Lemmas 5.1 and 5.3 in~\cite{lei2015}, we have the following lemma.
												
		\begin{lemma}\label{lma2}
			Let Assumption~\ref{asp1} hold. Suppose an $(1+\epsilon)$-approximate $k$-means algorithm is used in Algorithm~\ref{alg1} for a constant $\epsilon>0$. There exists a constant $c>0$ such that if
			\begin{equation*}
				\frac{Kd}{n_K\lambda_K^2\tilde{\theta}_{min}^2}<c,
			\end{equation*}
			then
			\begin{equation*}
				l(g,\hat{g}) =O\left(\frac{Kd}{\lambda_K^2\tilde{\theta}_{min}^2}\right)
			\end{equation*}
			with high probability.
		\end{lemma}
												
		Again we consider the $m$-uniform HPPM with Assumptions~\ref{asp1} and~\ref{asp2} being held. Algorithm~\ref{alg1} is weakly consistent when $K=O(n^{\frac{m-1}{2m-1}})$ and $\alpha_m=\omega(\frac{K^{2m-1}}{n^{m-1}})$, which is a weaker condition than that of \cite{ghoshdastidar2017}.
												
		\begin{remark}
			As reported in~\cite{ghoshdastidar2017}, for dense enough hypergraphs, one has $l(g,\hat{g})=O(1)$, which also implies strong consistency. Taking the $m$-uniform HPPM where Assumptions~\ref{asp1} and~\ref{asp2} hold for an example, one has $l(g,\hat{g})=O(1)$ whenever $m\geqslant 3$, $K=o(n^{\frac{m-2}{2m-2}})$ and $\alpha_m=\omega(\frac{K^{2m-2}}{n^{m-2}})$. In this case, it follows that $d=\omega(K^{2m-2}n)$, which means a much denser hypergraph than that implied by Theorem~\ref{thm1}.
		\end{remark}
												
		By the Davis-Kahan theorem, one can bound $\lVert \hat{U}^*\hat{O}-U^*\rVert_F$. However, it will lead to a trivial bound on $\lVert \hat{U}^*\hat{O}-U^*\rVert_{2,\infty}$, and thus could not guarantee strong consistency. As an alternative, we turn to more refined analysis of the perturbation of eigenvectors. For this purpose, we study the entry-wise eigenvector deviation between $\hat{U}$ and $U$. Due to the possibility of the presence of identical eigenvalues of $A$ ($P$), $\hat{U}$ ($U$) may not be uniquely determined. Therefore, a $K\times K$ orthogonal matrix is involved to align $\hat{U}$ and $U$.
												
		Let $H=\hat{U}^TU$ and its singular value decomposition be $H=\bar{U}\bar{\Sigma}\bar{V}^T$, then the matrix sign function~\cite{gross2011} is a $K\times K$ orthogonal matrix given by
		\begin{equation}\label{sgnmat}
			\mathrm{sgn}(H)=\bar{U}\bar{V}^T.
		\end{equation}
		We have the following theorem specifying an upper bound on $\lVert \hat{U}\mathrm{sgn}(H)-U\rVert_{2,\infty}$.
												
		\begin{theorem}\label{thm3}
			Let Assumptions~\ref{asp1} and~\ref{asp3} hold. For any $r>0$, there exist some constants $C_3=C_3(M,c_0,r)$ and $C_4=C_4(M,c_0,r)$ such that if $\sqrt{d\log n}\leqslant C_3|\lambda_K|$, then
			\begin{equation*}
				\lVert\hat{U}\mathrm{sgn}(H)-U\rVert_{2,\infty}\leqslant C_4\frac{\sqrt{d\log n}}{|\lambda_K|}\lVert U\rVert_{2,\infty}
			\end{equation*}
			with probability at least $1-O(n^{-r})$.
		\end{theorem}
												
		One can find the proof in Appendix~\ref{appC}, which is based on the leave-one-out technique used to study the entry-wise eigenvector deviation for graphs where the adjacency matrix has independent entries~\cite{abbe2020}. The main challenge here stems from the dependency across entries of $A$.
												
		To derive the bound on $\lVert\hat{U}\mathrm{sgn}(H)-U\rVert_{2,\infty}$, a key step is to bound
		\begin{equation}\label{eq321}
			\lVert (A-P)(\hat{U}H-U)\rVert_{2,\infty}=\max_{l\in[n]}\ \lVert (A-P)_{l\cdot}(\hat{U}H-U)\rVert,
		\end{equation}
		which is challenging because of the dependence between $A_{l\cdot}$ and $\hat{U}H$. We employ the leave-one-out method to tackle this problem.
												
		For each $l\in[n]$, we define a matrix $A^{(l)}\in\mathbb{R}^{n\times n}$ as
		\begin{equation}\label{matal}
			A^{(l)}_{ij}=\sum_{e\in E:a_{el}=0}\frac{a_{ei}(a_{ej}-\delta_{ij})}{|e|-1}h_e+\sum_{e\in E:a_{el}>0}\frac{a_{ei}(a_{ej}-\delta_{ij})}{|e|-1}\mathbb{E}[h_e],
		\end{equation}
		where all the randomness contributed by the possible hyperedges containing node $l$ is eliminated. Denote the eigenvalues of $A^{(l)}$ by $\{\hat{\lambda}_i^{(l)}\}_{i=1}^n$, which are arranged in decreasing order of absolute value. Let $\hat{U}^{(l)}$ be a matrix that contains the $K$ leading eigenvectors of $A^{(l)}$ as columns and let $H^{(l)}=(\hat{U}^{(l)})^TU$. We have the following observations:
		\begin{itemize}
			\item $\mathbb{E}[A^{(l)}]=P$ and $A^{(l)}$ concentrates around $P$. As a consequence, $\hat{U}^{(l)}$ and $H^{(l)}$ should be close to $\hat{U}$ and $H$, respectively;
			\item $A_{l\cdot}$ and $\hat{U}^{(l)}$ are independent, which implies the independence between $A_{l\cdot}$ and $\hat{U}^{(l)}H^{(l)}$;
			\item $A-A^{(l)}$ and $\hat{U}^{(l)}$ are independent.
		\end{itemize}
		Now we can bound \eqref{eq321} as
		\begin{equation*}
			\begin{aligned}
				  & \lVert (A-P)_{l\cdot}(\hat{U}H-U)\rVert                             \\
				  & \leqslant \lVert(A-P)_{l\cdot}(\hat{U}H-\hat{U}^{(l)}H^{(l)})\rVert
				+\lVert(A-P)_{l\cdot}(\hat{U}^{(l)}H^{(l)}-U)\rVert\\
				  & \leqslant \lVert A-P\rVert\lVert\hat{U}H-\hat{U}^{(l)}H^{(l)}\rVert
				+\lVert(A-P)_{l\cdot}(\hat{U}^{(l)}H^{(l)}-U)\rVert.
			\end{aligned}
		\end{equation*}
		Applying the Davis-Kahan theorem yields
		\begin{equation*}
			\lVert\hat{U}H-\hat{U}^{(l)}H^{(l)}\rVert\leqslant\frac{2\lVert(A-A^{(l)})\hat{U}^{(l)}\rVert}{|\lambda_K|}.
		\end{equation*}
		Therefore, we are able to apply the matrix Bernstein inequality \cite[Theorem 1.6.2]{tropp2015introduction} to bound $\lVert(A-A^{(l)})\hat{U}^{(l)}\rVert$ and $\lVert(A-P)_{l\cdot}(\hat{U}^{(l)}H^{(l)}-U)\rVert$. See the proof for more details.
												
		\emph{\textbf{Proof}} [Proof of Theorem \ref{thm1}]:
		By Theorem \ref{thm3}, there are constants $c_1,c_2>0$ such that if $\sqrt{d\log n}/|\lambda_K|\leqslant c_1$, then
		$$
		\lVert\hat{U}\mathrm{sgn}(H)-U\rVert_{2,\infty}\leqslant c_2\frac{\sqrt{d\log n}}{|\lambda_K|}\lVert U\rVert_{2,\infty}
		$$
		with probability at least $1-O(n^{-3})$. We first let $C_1,C_2<1/c_2$ and have
		$$\begin{aligned}
		\lVert\hat{U}_{i\cdot}\rVert&\geqslant\lVert U_{i\cdot}\rVert-\lVert\hat{U}_{i\cdot}\mathrm{sgn}(H)-U_{i\cdot}\rVert\\&\geqslant\min_{i\in[n]}\lVert U_{i\cdot}\rVert-\lVert\hat{U}\mathrm{sgn}(H)-U\rVert_{2,\infty}\\&\geqslant\min_{i\in[n]}\lVert U_{i\cdot}\rVert\left(1-c_2\frac{\gamma\sqrt{d\log n}}{|\lambda_K|}\right)>0,
		\end{aligned}$$
		with probability at least $1-O(n^{-3})$. For any two non-zero vectors $x,y\in\mathbb{R}^{K}$, we have
		$$
		\lVert \frac{x}{\lVert x\rVert}-\frac{y}{\lVert y\rVert}\rVert\leqslant2\frac{\lVert x-y\rVert}{\lVert y\rVert},
		$$
		which yields
		$$
		\begin{aligned}
			\lVert \hat{U}_{i\cdot}^*\mathrm{sgn}(H)-U_{i\cdot}^*\rVert & =\left\lVert \frac{\hat{U}_{i\cdot}}{\lVert\hat{U}_{i\cdot}\rVert}\mathrm{sgn}(H)-\frac{U_{i\cdot}}{\lVert U_{i\cdot}\rVert}\right\rVert \\&\leqslant2\frac{\lVert\hat{U}_{i\cdot}\mathrm{sgn}(H)-U_{i\cdot}\rVert}{\lVert U_{i\cdot}\rVert}\\&\leqslant2\frac{\lVert\hat{U}\mathrm{sgn}(H)-U\rVert_{2,\infty}}{\min_{i\in[n]}\lVert U_{i\cdot}\rVert}\\&\leqslant 2c_2\frac{\gamma\sqrt{d\log n}}{|\lambda_K|}.
		\end{aligned}
		$$
		Thus, we obtain
		$$\lVert \hat{U}^*\mathrm{sgn}(H)-U^*\rVert_{2,\infty}\leqslant 2c_2\frac{\gamma\sqrt{d\log n}}{|\lambda_K|}$$
		with probability exceeding $1-O(n^{-2})$.
												
		For strong consistency of the $k$-means step of Algorithm~\ref{alg1}, we use the result of Theorem 2.3 in~\cite{su2019}. Notice that the $K$ distinct rows of $U^*$ are orthogonal unit-length vectors, one can choose the two deterministic sequences as $c_{1n}=1$ and $c_{2n}=2c_2\frac{\gamma\sqrt{d\log n}}{|\lambda_K|}$. Since
		$$
		\sum_{n=1}^{\infty}\mathbb{P}(\lVert \hat{U}^*\mathrm{sgn}(H)-U^*\rVert_{2,\infty}\geqslant c_{2n})<\infty,
		$$
		then $\lVert \hat{U}^*\mathrm{sgn}(H)-U^*\rVert_{2,\infty}\leqslant c_{2n}$ almost surely. Under the current settings, a sufficient condition for Assumption 4.3 in~\cite{su2019} is
		$c_{2n}\leqslant c_3/(260K^{3/2})$, where the constant $c_3\in(0,1)$ is a lower bound on $Kn_K/n$ that does exist due to Assumption~\ref{asp2}. The result follows by choosing $C_1=\min\{c_1,c_3/(520c_2)\}$.
												
		Next, we consider the consistency of $\tilde{g}$. When nodes $i$ and $j$ belong to the same community, we have
		$$
		\begin{aligned}
			\lVert\hat{U}_{i\cdot}^*-\hat{U}_{j\cdot}^*\rVert & =\lVert(\hat{U}_{i\cdot}^*-\hat{U}_{j\cdot}^*)\mathrm{sgn}(H)\rVert                                                            \\
			                                                  & \leqslant\lVert\hat{U}_{i\cdot}^*\mathrm{sgn}(H)-U_{i\cdot}^*\rVert+\lVert\hat{U}_{j\cdot}^*\mathrm{sgn}(H)-U_{j\cdot}^*\rVert \\
			                                                  & \leqslant 4c_2\frac{\gamma\sqrt{d\log n}}{|\lambda_K|}.
		\end{aligned}
		$$
		While they belong to different communities, it follows that
		$$
		\begin{aligned}
			\lVert\hat{U}_{i\cdot}^*-\hat{U}_{j\cdot}^*\rVert & =\lVert(\hat{U}_{i\cdot}^*-\hat{U}_{j\cdot}^*)\mathrm{sgn}(H)\rVert                                                                                                   \\
			                                                  & \geqslant\lVert U_{i\cdot}^*-U_{j\cdot}^*\rVert-\lVert\hat{U}_{i\cdot}^*\mathrm{sgn}(H)-U_{i\cdot}^*\rVert-\lVert\hat{U}_{j\cdot}^*\mathrm{sgn}(H)-U_{j\cdot}^*\rVert
			\\&\geqslant\sqrt{2}- 4c_2\frac{\gamma\sqrt{d\log n}}{|\lambda_K|}.
		\end{aligned}
		$$
		Therefore, Algorithm~\ref{alg2} exactly recovers the true community structure whenever
		\begin{equation*}
			4c_2\frac{\gamma\sqrt{d\log n}}{|\lambda_K|}<1/\sqrt{2} \text{ and} \sqrt{2}-4c_2\frac{\gamma\sqrt{d\log n}}{|\lambda_K|}\geqslant1/\sqrt{2},
		\end{equation*}
		which are satisfied if $C_2<\min\{c_1,1/(4\sqrt{2}c_2)\}$.
		$\hfill\blacksquare$
												
		\section{Consistency for the DCHPPM}\label{specialcases}
												
		\subsection{The Uniform Case}\label{uniform}
												
		\begin{corollary}\label{coro2}
			Let $A$ be a weighted adjacency matrix of a hypergraph generated by an $m$-uniform DCHPPM, then Assumption~\ref{asp4} holds. If
			\begin{equation*}
				K^{m-1}{\phi_{max}^2}/{\phi_{min}^2}=O(\sqrt{\log n}),
			\end{equation*}
			then Assumption~\ref{asp3} holds. Suppose Assumptions~\ref{asp1}-\ref{asp4} hold.
			\begin{itemize}
				\item[(i).]
				      There exists a constant $C_1>0$ such that for sufficiently large $n$, if
				      \begin{equation}
				      	\alpha_m\geqslant C_1\frac{\gamma^2\theta_{max}^2K^{2m+1}\log n}{n^{m-1}},
				      \end{equation}
				      then $\hat{g}=g$ with high probability.
				\item[(ii).]
				      There exists a constant $C_2>0$ such that if
				      \begin{equation}
				      	\alpha_m\geqslant C_2\frac{\gamma^2\theta_{max}^2K^{2m-2}\log n}{n^{m-1}},
				      \end{equation}
				      then $\tilde{g}=g$ with high probability.
			\end{itemize}
		\end{corollary}
												
		\subsection{The Non-uniform Case}\label{nonuniform}
												
		\begin{corollary}\label{coro3}
			Let $A$ be a weighted adjacency matrix of a hypergraph generated by the above model, then Assumption~\ref{asp4} holds. Let $m_0=\max\{m\in\mathbb{Z}|2\leqslant m\leqslant M,\alpha_m>0\}$. If
			\begin{equation*}
				K^{m_0-1}{\phi_{max}^2}/{\phi_{min}^2}=O(\sqrt{\log n}),
			\end{equation*}
			then Assumption~\ref{asp3} holds. Suppose Assumptions~\ref{asp1}-\ref{asp4} hold.
			\begin{itemize}
				\item[(i).]
				      There exists a constant $C_1>0$ such that for sufficiently large $n$, if
				      \begin{equation}\label{coro3condition1}
				      	\sum_{m=2}^Mm\alpha_mn^{m-1}\geqslant C_1{\gamma^2\theta_{max}^2K^{2m_0+1}\log n},
				      \end{equation}\label{coro3condition2}
				      then $\hat{g}=g$ with high probability.
				\item[(ii).]
				      There exists a constant $C_2>0$ such that if
				      \begin{equation}
				      	\sum_{m=2}^Mm\alpha_mn^{m-1}\geqslant C_2{\gamma^2\theta_{max}^2K^{2m_0-2}\log n},
				      \end{equation}
				      then $\tilde{g}=g$ with high probability.
			\end{itemize}
		\end{corollary}
												
		\begin{remark}
			As the number of hyperedges of size $m$ is of order $\Theta(\alpha_mn^{m})$, \eqref{coro3condition1} essentially requires that the total number of hyperedges is $\Omega(\gamma^2\theta_{max}^2K^{2m_0+1}n\log n)$, which reduces to $\Omega(n\log n)$ when $\theta_i=1$ for all $i\in[n]$ and $K$ is a constant.
		\end{remark}
												
		\section{Numerical experiment}\label{experiment}
												
		In this section, we test the practical performance of Algorithms 1 and 2 in DCHPPMs. We set $n=3000, M=4, p = 10, q = 1$ and vary $K,\theta,$ and the averaged expected degree. We choose $K$ from $\{2,3\}$ and let the communities have the same size in both cases. We consider two choices of $\theta$: (i) $\theta_i^{(1)}=1$ for all $i\in[n]$; and (ii) first draw $\psi_i$ uniformly and independently from $[1,2]$, and then set $\theta_i^{(2)}=n\psi_i/(K\sum_{j\in[n]}\psi_i\delta_{g_i,g_j})$. We choose $\alpha_m$ such that the number of hyperedges of different sizes is the same. By multiplying all $\alpha_m$ by the same number, we are able to vary sparsity of the hypergraph. In implementation of Algorithm 1, we use MATLAB \lq\lq kmeans\rq\rq~algorithm.
												
\begin{figure}[htbp]
			\centering
			\includegraphics[width=0.48\textwidth]{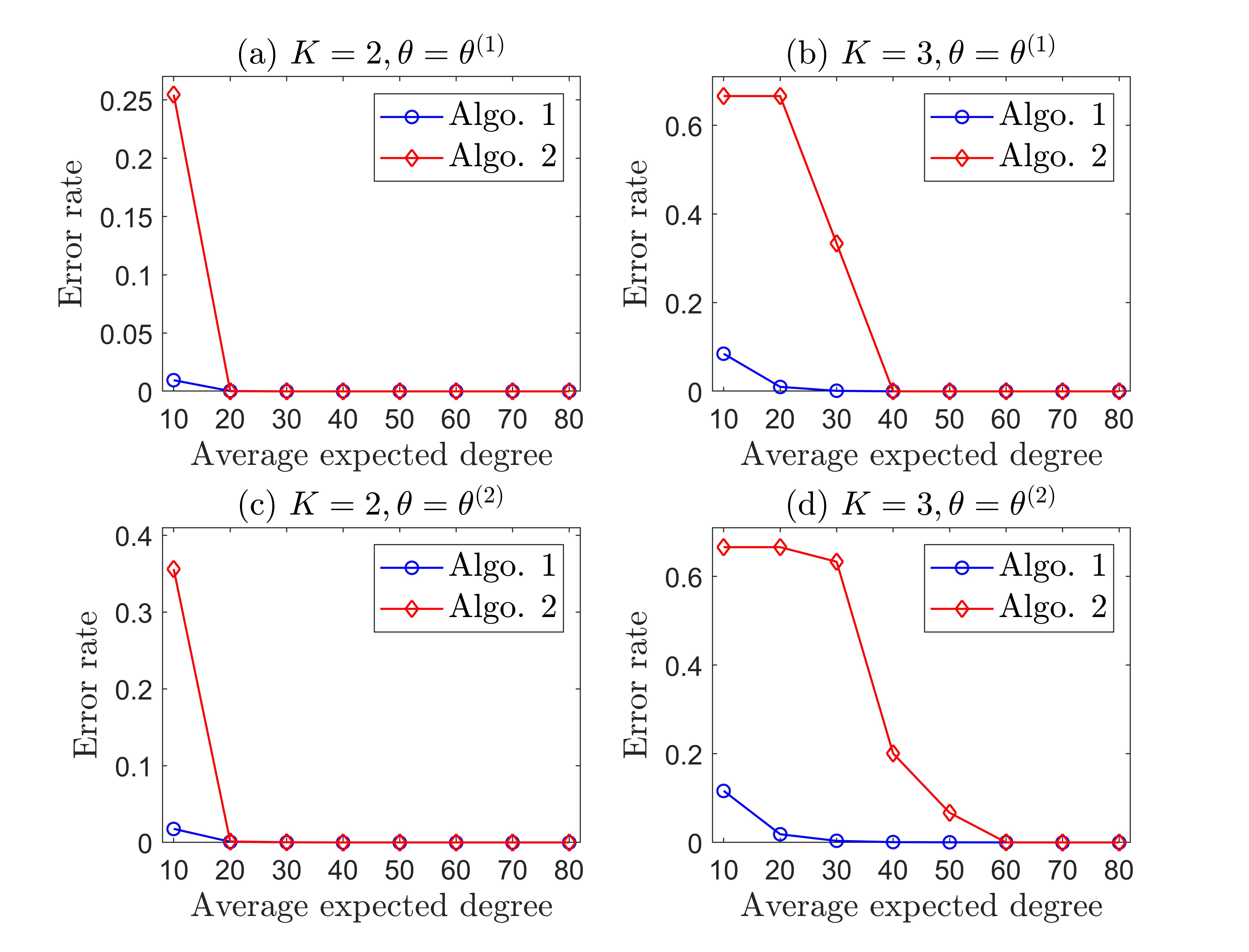}
			\caption{Clustering result for Algorithms 1 and 2 in the DCHPPM. Each value is the average over 10 independent trials. The first and second rows show the results corresponding to $\theta=\theta^{(1)}$ and $\theta^{(2)}$, respectively. The first and second columns are the results corresponding to $K=2$ and 3, respectively.}
			\label{fig1}
		\end{figure}

		The clustering accuracy measured by an error rate $l(g,g')/n$ is shown in Figure \ref{fig1}. For both Algorithms \ref{alg1} and \ref{alg2}, larger number of communities or stronger degree heterogeneity will make it harder to achieve exact recovery. As the hypergraph gets denser, both algorithms are able to fully recover the true community assignment. In the experiment, Algorithm 1 consistently outperforms Algorithms 2, which demonstrates the superiority of the $k$-means algorithm over the simple thresholding method.
												
		\section{Conclusion}\label{conclusion}
		This paper characterizes the performance of spectral clustering based on the weighted adjacency matrix in a sparse DCHSBM.
		We consider both heterogeneous hyperdegrees and general edge-connecting probability.
		By establishing an entry-wise eigenvector perturbation bound, we show that even the simplest spectral clustering algorithm could exactly recover the true community structure under mild conditions on the model parameters, hence an important advancement over the literature.
		We suggest two further extensions. As spectral clustering assumes that the number of communities $K$ is known, an important question would be the estimation of $K$.
		In this paper, we allow $K$ to grow with $n$, but at a very slow rate to ensure that Assumption~\ref{asp3} holds.
		One possible way to improve this is to use the method developed in Ref.~\cite{mao2021} to establish the entry-wise bound.
		However, this may fail when the maximum expected degree grows as $\Omega(\log n)$, the most challenging regime.
		On the other hand, a recent study~\cite{zhang2021} characterized the theoretical limit for exact recovery in a general uniform HSBM by the GCH-divergence.
		While the success of hypergraph spectral clustering depends on a large eigen-gap, it is important to explore and interpret the relation between $\lambda_K$ and the GCH-divergence.
		The proposed spectral algorithm, relying on local refinement to be strongly consistent, is shown to meet the exact recovery threshold.
		It is worth investigating whether the spectral algorithms considered in this work achieve the theoretical limit.
		If not, can the threshold be achieved by adding a local refinement step~\cite{zhang2021}?
		Since hypergraph projection itself may incur information loss which makes it hard to achieve the limit, a possible approach is to consider tensor-based methods.

												
		%

		\appendices
		\section{Proof of lemmas and corollaries}\label{appA}
												
		\subsection{Proof of Lemma~\ref{lma1}}\label{prooflm1}
												
		(i) For $i\neq j$, consider a hyperedge $e$ such that $a_{ei}\geqslant 1$ and $a_{ej}\geqslant 1$, one has $b_e=\frac{|e|!}{\prod_{k=1}^na_{ek}!}$ and $b_{e\backslash\{i,j\}}=\frac{(|e|-2)!}{(\prod_{k\neq i,j}a_{ek}!)\cdot(a_{ei}-1)!\cdot(a_{ej}-1)!}=\frac{a_{ei}a_{ej}}{|e|(|e|-1)}b_e$, and thus
		$$A_{ij}=\sum_{e\in E:a_{ei}\geqslant1,a_{ej}\geqslant1}|e|\frac{b_{e\backslash\{i,j\}}}{b_e}h_e.$$
		Then
		$$\begin{aligned}
		&P_{ij}=\sum_{e\in E:a_{ei}\geqslant1,a_{ej}\geqslant1}|e|\frac{b_{e\backslash\{i,j\}}}{b_e}\mathbb{E}[h_e]\\
		&=\sum_{m=2}^Mm\sum_{i_3=1}^n\cdots\sum_{i_m=1}^n\theta_i\theta_j\prod_{l=3}^m\theta_{i_l}\cdot\Phi(g_i,g_j,g_{i_3},\cdots,g_{i_m})\\
		&=\theta_i\theta_j\sum_{m=2}^Mm\sum_{k_3=1}^K\cdots\sum_{k_m=1}^K\prod_{l=3}^mn_{k_l}\cdot\Phi(g_i,g_j,k_3,\cdots,k_m).
		\end{aligned}$$
												
		For each $e\in E$ such that $a_{ei}\geqslant2$, one obtains $b_{e\backslash\{i,i\}}=\frac{(|e|-2)!}{(\prod_{j\neq i}a_{ej}!)\cdot(a_{ei}-2)!}=\frac{a_{ei}(a_{ei}-1)}{|e|(|e|-1)}b_e$. Therefore
		$$A_{ii}=\sum_{e\in E:a_{ei}\geqslant2}|e|\frac{b_{e\backslash\{i,i\}}}{b_e}h_e,$$
		and
		$$\begin{aligned}
		&P_{ii}=\sum_{e\in E:a_{ei}\geqslant2}|e|\frac{b_{e\backslash\{i,i\}}}{b_e}\mathbb{E}[h_e]\\
		&=\sum_{m=2}^Mm\sum_{i_3=1}^n\cdots\sum_{i_m=1}^n\theta_i^2\prod_{l=3}^m\theta_{i_l}\cdot\Phi(g_i,g_i,g_{i_3},\cdots,g_{i_m}){}\\
		&=\theta_i^2\sum_{m=2}^Mm\sum_{k_3=1}^K\cdots\sum_{k_m=1}^K\prod_{l=3}^mn_{k_l}\cdot\Phi(g_i,g_i,k_3,\cdots,k_m).
		\end{aligned}$$
												
		The result follows by letting $B$ be a $K\times K$ matrix with
		$$B_{rs}=\sum_{m=2}^Mm\sum_{k_3=1}^K\cdots\sum_{k_m=1}^K\prod_{l=3}^mn_{k_l}\cdot\Phi(r,s,k_3,\cdots,k_m),$$
		for $1\leqslant r,s\leqslant K$.
												
		(ii) See, for example, Lemma 4.1 in~\cite{lei2015}.
												
		\subsection{Proof of Lemma~\ref{lma2}}\label{prooflm2}
												
		By Lemma 5.1 in \cite{lei2015}, there exists an orthogonal matrix $\hat{Q}\in\mathbb{R}^{K\times K}$ such that
		$$\lVert\hat{U}\hat{Q}-U\rVert_F\leqslant\frac{2\sqrt{2K}}{|\lambda_K|}\lVert A-P\rVert\leqslant C\frac{2\sqrt{2Kd}}{|\lambda_K|}$$
		with high probability.
												
		Note that one cannot guarantee $\lVert\hat{U}_{i\cdot}\rVert>0$ for all $i\in[n]$ with high probability. Let $S_0=\{i\in[n]|\hat{U}_{i\cdot}=0\}$ be the set of nodes that correspond to zero rows of $\hat{U}$. The nodes in $S_0$ are regarded as mis-clustered and are not involved in the $k$-means step. Since $\lVert\hat{U}\hat{Q}-U\rVert_F^2\geqslant\sum_{i=1}^n\mathbbm{1}_{\{\hat{U}_{i\cdot}=0\}}\lVert U_{i\cdot}\rVert^2\geqslant\tilde{\theta}_{min}^2|S_0|$, we have $|S_0|\leqslant\lVert\hat{U}\hat{Q}-U\rVert_F^2/\tilde{\theta}_{min}^2$.
												
		Let $\hat{U}'$ and $U'$ be the matrices whose rows are given by the normalized non-zero rows of $\hat{U}$ and $U$, respectively, it follows that
		$$\begin{aligned}
		\lVert\hat{U}'\hat{Q}-U'\rVert_F&=\sqrt{\sum_{i\in S_0^c}\left\lVert\frac{\hat{U}_{i\cdot}}{\lVert\hat{U}_{i\cdot}\rVert}\hat{Q}-\frac{U_{i\cdot}}{\lVert U_{i\cdot}\rVert}\right\rVert^2}\\
		&\leqslant\sqrt{\sum_{i=1}^n\frac{\lVert\hat{U}_{i\cdot}\hat{Q}-U_{i\cdot}\rVert^2}{\lVert U_{i\cdot}\rVert^2}}\\
		&\leqslant \frac{\lVert\hat{U}\hat{Q}-U\rVert_F}{\tilde{\theta}_{min}}.
		\end{aligned}$$
		Define $S_1=\{i\in S_0^c|\lVert \hat{U}_{i\cdot}'\hat{Q}-U_{i\cdot}'\rVert\geqslant 1/\sqrt{2}\}$, then
		$$|S_1|\leqslant4(2+\epsilon)\lVert\hat{U}'\hat{Q}-U'\rVert_F^2$$
		By Lemma 5.3 in \cite{lei2015}. When
		$$n_K>4(2+\epsilon)\lVert\hat{U}'\hat{Q}-U'\rVert_F^2,$$
		the nodes outside $S_0\bigcup S_1$ are all correctly assigned.
												
		Thus, one can choose $c=\frac{1}{32(2+\epsilon)C}$ so that when
		$$\frac{Kd}{n_K\lambda_K^2\tilde{\theta}_{min}^2}<c,$$
		the nodes outside $S_0\bigcup S_1$ are all correctly clustered, and the number of incorrectly clustered nodes is bounded by
		$$l(g,\hat{g})\leqslant |S_0|+|S_1|\leqslant8(9+4\epsilon)C\frac{Kd}{\lambda_K^2\tilde{\theta}_{min}^2}.$$
												
		\subsection{Proof of Corollary~\ref{coro2}}\label{proofcoro2}
												
		Under the current settings, it is easy to obtain
		$$B_{rs}=m\alpha_{m}\left((p-q)n_r^{m-2}\delta_{rs}+qn^{m-2}\right).$$
		Notice that the eigenvalues of $P$ are identical to those of $\mathrm{diag}(\phi) B\mathrm{diag}(\phi)$, we have
		$$\begin{aligned}
		&\lambda_{max}(B)\leqslant m\alpha_m((p-q)n_1^{m-2}+Kqn^{m-2})\\
		&~~~~~~~~~~~\leqslant Kpm\alpha_mn^{m-2},\\
		&\lambda_1\leqslant\lambda_{max}(B)\cdot\max_{k}\phi_k^2\leqslant Kpm\alpha_m\phi_{max}^2n^{m-2},\\
		&\lambda_{min}(B)\geqslant (p-q)m\alpha_{m}n_{K}^{m-2},\\
		&\lambda_K\geqslant\lambda_{min}(B)\cdot{\min_{k}\phi_k^2}\geqslant (p-q)m\alpha_m\phi_{min}^2n_K^{m-2},
		\end{aligned}$$
		and then $\kappa=\lambda_1/\lambda_K\leqslant c_1K^{m-1}\phi_{max}^2/\phi_{min}^2$ for some constant $c_1>0$. Since $\phi_k^2\geqslant (\sum_{i}\theta_i\delta_{g_i,k})^2/n_k=n_k$, we have $\lambda_K\geqslant (p-q)m\alpha_mn_K^{m-1}$. When $\max_{ij}P_{ij}\leqslant\theta_{max}^2\max_{rs}B_{rs}\leqslant pm\alpha_m\theta_{max}^2n^{m-2}$, we have $d\leqslant\max\{ pm\alpha_m\theta_{max}^2n^{m-1},c_0\log n\}$.
												
		Consider the consistency of Algorithm~\ref{alg1}, a sufficient condition for $\gamma K^{3/2}\frac{\sqrt{d\log n}}{|\lambda_K|}$ being sufficiently small is
		$$\alpha_m\geqslant C_1\frac{\gamma^2\theta_{max}^2K^{2m+1}\log n}{n^{m-1}}$$
		for some constant $C_1>0$.
												
		Analogously, there exists a constant $C_2>0$ such that if
		$$\alpha_m\geqslant C_2\frac{\gamma^2\theta_{max}^2K^{2m-2}\log n}{n^{m-1}},$$
		then Algorithm~\ref{alg2} is strongly consistent.
												
		\subsection{Proof of Corollary~\ref{coro3}}\label{proofcoro3}
												
		It is easy to obtain the following conclusions:
		$$\begin{aligned}
		&B_{rs}=\sum_{m=2}^Mm\alpha_{m}\left((p-q)n_r^{m-2}\delta_{rs}+qn^{m-2}\right),\\
		&\lambda_{max}(B)\leqslant Kp\sum_{m=2}^Mm\alpha_mn^{m-2},\\
		&\lambda_{min}(B)\geqslant(p-q)\sum_{m=2}^Mm\alpha_mn_K^{m-2}\geqslant c_1\frac{p-q}{K^{m_0-2}}\sum_{m=2}^Mm\alpha_mn^{m-2},\\
		&\lambda_1\leqslant \phi_{max}^2\lambda_{max}(B)\leqslant Kp\phi_{max}^2\sum_{m=2}^Mm\alpha_mn^{m-2},\\
		&\lambda_K\geqslant \phi_{min}^2\lambda_{min}(B)\geqslant \frac{c_1(p-q)\phi_{min}^2}{K^{m_0-2}}\sum_{m=2}^Mm\alpha_mn^{m-2},\\
		&\kappa\leqslant \frac{p}{c_1(p-q)}K^{m_0-1}\phi_{max}^2/\phi_{min}^2,
		\end{aligned}$$
		where $c_1>0$ is a constant.
												
		From the fact that $\phi_{min}\geqslant\sqrt{n_K}$, we have $\lambda_K\geqslant\frac{c_2(p-q)}{K^{m_0-1}}\sum_{m=2}^Mm\alpha_mn^{m-1}$ for some constant $ c_2>0$.
		Since $\max_{ij}P_{ij}\leqslant p\theta_{max}^2\sum_{m=2}^Mm\alpha_mn^{m-2}$, we have $d\leqslant \max\{p\theta_{max}^2\sum_{m=2}^Mm\alpha_mn^{m-1},c_0\log n\}$. Then, the sufficient conditions for ${\gamma K^{3/2}\sqrt{d\log n}}/{|\lambda_K|}$ and ${\gamma\sqrt{d\log n}}/{|\lambda_K|}$ being sufficiently small are
		$$\sum_{m=2}^Mm\alpha_mn^{m-1}\geqslant C_1\gamma^2\theta_{max}^2K^{2m_0+1}\log n$$
		for some constant $C_1>0$ and
		$$\sum_{m=2}^Mm\alpha_mn^{m-1}\geqslant C_2\gamma^2\theta_{max}^2K^{2m_0-2}\log n$$
		for some constant $C_2>0$, respectively.
												
		\section{Proof of Theorem~\ref{thm2}}\label{appB}
												
		The overall structure of the proof is similar to that in \cite{lei2015}. Let $W=A-P$ and $S=\{x\in\mathbb{R}^n|\lVert x\rVert\leqslant1\}$ be the unit ball in $\mathbb{R}^n$, the goal is to bound the spectral norm of $W$:
		$$\lVert W\rVert=\sup_{x\in S}|x^TWx|.$$
		We first outline the three major steps of the proof.
												
		\begin{itemize}
			\item[(1).] Discretization. Fix a constant $\delta\in(0,1/3)$ and define a set of grid points in $S$:
			      $$T=\{x=(x_1,\cdots,x_n)^T\in S|\sqrt{n}x_i/\delta\in\mathbb{Z},\text{for }i\in[n]\},$$
			      we will prove the following results:
			      \begin{itemize}
			      	\item[(i)] $T$ is a $\delta$-net of $S$. That is, $T$ is a finite subset of $S$ such that for any $x\in S$, there exists a point $y\in T$ satisfying $\lVert x-y\rVert\leqslant \delta$.
			      	\item[(ii)] $\lVert W\rVert\leqslant(1-3\delta)^{-1}\sup_{x,y\in T}|x^TWy|.$
			      \end{itemize}
			      			      			      			      			      			
			      Bounding $\lVert W\rVert$ is then reduced to bound the supremum of $|x^TWy|$ over all $x,y\in T$. Next, we split the point pairs in the grid into two parts, called the \emph{light pairs} and \emph{heavy pairs}, and then bound each of the two parts separately.
			      			      			      			      			      			
			\item[(2).] Bounding the light pairs. We use Bernstein inequality to bound the contribution of the light pairs.
			      			      			      			      			      			
			\item[(3).] The contribution of heavy pairs, however, cannot be simply bounded by the standard Bernstein inequality. We show that the two key properties of the random hypergraph, i.e., \emph{bounded degree property} and \emph{bounded  discrepancy property}, still hold and thus the contribution of heavy pairs is bounded.
		\end{itemize}
												
		\begin{lemma}\label{thm2lm1}
			$T$ is a $\delta$-net of $S$.
		\end{lemma}
												
		\emph{Proof}: It is straightforward to see that $T$ is a finite subset of $S$. For any $x=(x_1,\cdots,x_n)^T\in S$, without loss of generality, assume $x_i\geqslant0$ for $i=1,\cdots,n$. For each $1\leqslant i \leqslant n$, there exists $k_i\in\mathbb{Z}$ such that
		$$0\leqslant k_i\delta/\sqrt{n}\leqslant x_i\leqslant (k_i+1)\delta/\sqrt{n}.$$
		Let $y=(k_1\delta/\sqrt{n},\cdots,k_n\delta/\sqrt{n})^T$, we have $\lVert y\rVert\leqslant\lVert x\rVert\leqslant 1,$ which means $y\in T$. Moreover, we have $\lVert x-y\rVert\leqslant \sqrt{n(\delta/\sqrt{n})^2}=\delta$.
												
		\begin{lemma}\label{thm2lm2}
			$\lVert W\rVert\leqslant(1-3\delta)^{-1}\sup_{x,y\in T}|x^TWy|.$
		\end{lemma}
												
		\emph{Proof}: For any $x_0\in S$, according to Lemma \ref{thm2lm1}, there exist $x_1,x_2\in T$ such that $\lVert x_0-x_1\rVert\leqslant\delta$ and $\lVert x_0-x_2\rVert\leqslant\delta$. We have
		$$\begin{aligned}
		|x_0^TWx_0|&=|(x_0-x_1+x_1)^TW(x_0-x_2+x_2)|\\
		&\leqslant|(x_0-x_1)^TW(x_0-x_2)|+|(x_0-x_1)^TWx_2|\\
		&~~~+|x_1^TW(x_0-x_2)|+|x_1^TWx_2|\\
		&\leqslant(\delta^2+2\delta)\lVert W\rVert+|x_1^TWx_2|\\&\leqslant3\delta\lVert W\rVert+\sup_{x,y\in T}|x^TWy|.
		\end{aligned}$$
		Then $\lVert W\rVert=\sup_{x\in S}|x^TWx|\leqslant 3\delta\lVert W\rVert+\sup_{x,y\in T}|x^TWy|.$
												
		Split the pairs $(x_i,y_j)$ into light pairs: $L_{xy}=\{(i,j)||x_iy_j|\leqslant\sqrt{d}/n\}$ and heavy pairs $\bar{L}_{xy}=\{(i,j)||x_iy_j|>\sqrt{d}/n\}$. The corresponding contributions are $I_{xy}=\sum_{(i,j)\in L_{xy}}x_iy_jw_{ij}$ and $\bar{I}_{xy}=\sum_{(i,j)\in \bar{L}_{xy}}x_iy_jw_{ij}.$
												
		\begin{lemma}\label{thm2lm3}
			For any constant $c>0$,
			$$\begin{aligned}
			&\mathbb{P}\left(\sup_{x,y\in T}|I_{xy}|\geqslant c\sqrt{d}\right)\\
			&\leqslant 2\exp\left[-\left(\frac{c^2/2}{M(M^2+c/3)}-2\log\left(\frac{2}{\delta}+1\right)\right)n\right].
			\end{aligned}$$
		\end{lemma}
												
		\emph{Proof}: For any $x,y\in T$, we have
		$$\begin{aligned}
		I_{xy}&=\sum_{(i,j)\in L_{xy}}x_iy_j\sum_{e\in E}\frac{a_{ei}(a_{ej}-\delta_{ij})}{|e|-1}(h_e-\mathbb{E}[h_e])\\&=\sum_{e\in E}\left(\sum_{(i,j)\in L_{xy}}\frac{a_{ei}(a_{ej}-\delta_{ij})x_iy_j}{|e|-1}\right)(h_e-\mathbb{E}[h_e]).
		\end{aligned}$$
												
		Let $S_e=\left(\sum_{(i,j)\in L_{xy}}\frac{a_{ei}(a_{ej}-\delta_{ij})x_iy_j}{|e|-1}\right)(h_e-\mathbb{E}[h_e])$, then $I_{xy}=\sum_{e\in E}S_e$. Note that $\{S_e\}_{e\in E}$ are independent random variables, and
												
		$$\mathbb{E}[S_e]=0,$$
		$$|S_e|\leqslant \frac{1}{|e|-1}\cdot\frac{\sqrt{d}}{n}\sum_{(i,j)\in L_{xy}}a_{ei}(a_{ej}-\delta_{ij})\leqslant M\frac{\sqrt{d}}{n},$$
		$$\begin{aligned}
		\sum_{e\in E}\mathbb{E}[S_e^2]&=\sum_{e\in E}\frac{var(h_e)}{(|e|-1)^2}\left(\sum_{(i,j)\in L_{xy}}a_{ei}(a_{ej}-\delta_{ij})x_iy_j\right)^2\\
		&\leqslant\sum_{e\in E}\mathbb{E}[h_e]\frac{|e|}{|e|-1}\sum_{(i,j)\in L_{xy}}a_{ei}^2(a_{ej}-\delta_{ij})^2x_i^2y_j^2\\
		&\leqslant \sum_{e\in E}{\mathbb{E}[h_e]}\frac{|e|}{|e|-1}\sum_{1\leqslant i,j\leqslant n}a_{ei}^2(a_{ej}-\delta_{ij})^2x_i^2y_j^2\\
		&\leqslant \sum_{e\in E}{\mathbb{E}[h_e]}{|e|^2}\cdot\sum_{1\leqslant i,j\leqslant n}a_{ei}(a_{ej}-\delta_{ij})x_i^2y_j^2\\
		&\leqslant M^3\sum_{e\in E}{\mathbb{E}[h_e]}\cdot\sum_{1\leqslant i,j\leqslant n}\frac{a_{ei}(a_{ej}-\delta_{ij})x_i^2y_j^2}{|e|-1}\\
		&=M^3\sum_{1\leqslant i,j\leqslant n}\left(\sum_{e\in E}\frac{a_{ei}(a_{ej}-\delta_{ij})}{|e|-1}\mathbb{E}[h_e]\right)x_i^2y_j^2\\
		&=M^3\sum_{1\leqslant i,j\leqslant n}P_{ij}x_i^2y_j^2\\&\leqslant {M^3d}/{n}.
		\end{aligned}$$
												
		Applying the Bernstein inequality yields
		$$\begin{aligned}
		\mathbb{P}(|I_{xy}|\geqslant c\sqrt{d})&=\mathbb{P}(|\sum_{e\in E}S_e|\geqslant c\sqrt{d})\\&\leqslant2\exp\left(\frac{-c^2d/2}{\frac{M^3d}{n}+M\frac{\sqrt{d}}{n}\cdot c\sqrt{d}/3}\right)\\&=2\exp\left(-\frac{c^2/2}{M(M^2+c/3)}n\right).
		\end{aligned}$$
		Since $|T|\leqslant (\frac{2}{\delta}+1)^n$, we have
		$$\begin{aligned}
		&\mathbb{P}\left(\sup_{x,y\in T}|I_{xy}|\geqslant c\sqrt{d}\right)\leqslant |T|^2\cdot2\exp\left(-\frac{c^2/2}{M(M^2+c/3)}n\right)\\
		&\leqslant 2\exp\left[-\left(\frac{c^2/2}{M(M^2+c/3)}-2\log\left(\frac{2}{\delta}+1\right)\right)n\right].
		\end{aligned}$$
												
		Next, we show that $\sup_{x,y\in T}|\bar{I}_{xy}|$ is bounded by $c\sqrt{d}$. Since for any $x,y\in T$,
		$$\begin{aligned}
		|\sum_{(i,j)\in\bar{L}_{xy}}x_iy_jP_{ij}|&\leqslant \sum_{(i,j)\in\bar{L}_{xy}}\frac{x_i^2y_j^2}{|x_iy_j|}P_{ij}\\
		&\leqslant\frac{n}{\sqrt{d}}p_{max}\sum_{1\leqslant i,j\leqslant n}x_i^2y_j^2\\
		&\leqslant \sqrt{d},
		\end{aligned}$$
		we need only to show that
		$$|\sum_{(i,j)\in\bar{L}_{xy}}x_iy_jA_{ij}|=O(\sqrt{d})$$
		for any $x,y\in T$ (not with high probability but definitely). In what follows, we prove that both the bounded degree property and the bounded discrepancy property hold.
												
		\begin{lemma}[Bounded degree property]\label{thm2lm4}
			For any constant $c>0$, there exists a constant $c_1=c_1(c)$ such that $\mathbb{P}(d_i\geqslant c_1d)\leqslant n^{-c}$ for any node $i\in[n]$.
		\end{lemma}
												
		\emph{Proof}: Recall that $d_i=\sum_{j=1}^nA_{ij}=\sum_{e\in E}a_{ei}h_e.$ Notice that $\mathbb{E}[d_i]=\sum_{j=1}^n\mathbb{E}[A_{ij}]=\sum_{j=1}^nP_{ij}\leqslant d$, we have
		$$\begin{aligned}
		\mathbb{P}(d_i\geqslant c_1d)&=\mathbb{P}(d_i-\mathbb{E}[d_i]\geqslant c_1d-\mathbb{E}[d_i])\\&\leqslant \mathbb{P}(d_i-\mathbb{E}[d_i]\geqslant (c_1-1)d)\\&=\mathbb{P}\left(\sum_{e\in E}a_{ei}(h_e-\mathbb{E}[h_e])\geqslant(c_1-1)d\right).
		\end{aligned}$$
		Let $S_e=a_{ei}(h_e-\mathbb{E}[h_e])$. Since $\mathbb{E}[S_e]=0$, $|S_e|\leqslant M$, and
		$$\begin{aligned}
		\sum_{e\in E}\mathbb{E}[S_e^2]&=\sum_{e\in E}a_{ei}^2var(h_e)\\
		&\leqslant M\sum_{e\in E}a_{ei}\mathbb{E}[h_e]\\
		&=M\cdot\mathbb{E}[d_i]\\
		&\leqslant Md,
		\end{aligned}$$
		we have
		$$\begin{aligned}
		\mathbb{P}(d_i\geqslant c_1d)&\leqslant \exp\left(\frac{-(c_1-1)^2d^2/2}{Md+M\cdot(c_1-1)d/3}\right)\\
		&\leqslant\exp\left(-\frac{c_0(c_1-1)^2/2}{M(2+c_1)/3}\log n\right)\\
		&=n^{-\frac{3c_0(c_1-1)^2}{2M(2+c_1)}}
		\end{aligned}$$
		by the Bernstein inequality.
												
		For any $I,J\subset[n]$, let $e(I,J)=\sum_{i\in I}\sum_{j\in J}A_{ij}$ and define $\mu(I,J)=|I||J|\frac{d}{n}$ which is an upper bound on the expectation of $e(I,J)$.
		\begin{lemma}[Bounded discrepancy property]\label{thm2lm5}
			For any constant $c>0$, there exist constants $c_2=c_2(c)$ and $c_3=c_3(c)$, such that, with probability least $1-2n^{-c}$, for any $I,J\subset[n]$ with $|I|\leqslant |J|$, either of the following holds:
			\begin{itemize}
				\item $e(I,J)/\mu(I,J)\leqslant \mathrm{e} c_2$ (here $\mathrm{e}$ denotes Euler's number).
				\item $e(I,J)\log\frac{e(I,J)}{\mu(I,J)}\leqslant c_3|J|\log\frac{n}{|J|}$.
			\end{itemize}
		\end{lemma}
												
		\emph{Proof}: If $|J|\geqslant n/\mathrm{e}$, according to Lemma~\ref{thm2lm4}, there exists $c_1=c_1(c)$ such that $e(I,J)\leqslant\sum_{i\in I}d_i\leqslant c_1d|I|$ with probability at least $1-n^{-c}$. Since $\mu(I,J)\geqslant \frac{d}{\mathrm{e}}|I|$, we have $e(I,J)/\mu(I,J)\leqslant \mathrm{e} c_1$ with high probability.
												
		Otherwise, suppose $|J|< n/\mathrm{e}$. Note that
		$$\begin{aligned}
		e(I,J)&=\sum_{i\in I}\sum_{j\in J}\sum_{e\in E}\frac{a_{ei}(a_{ej}-\delta_{ij})}{|e|-1}h_e\\
		&=\sum_{e\in E}\left(\sum_{i\in I,j\in J}\frac{a_{ei}(a_{ej}-\delta_{ij})}{|e|-1}\right)h_e
		\end{aligned}$$
		is a sum over independent random variables and
		$$0\leqslant \left(\sum_{i\in I,j\in J}\frac{a_{ei}(a_{ej}-\delta_{ij})}{|e|-1}\right)h_e\leqslant |e|\leqslant M.$$ According to Lemma 5 in~\cite{ahn2018}, there exists a constant $c'=c'(M)>1$ such that
		$$\mathbb{P}(e(I,J)>k\mu(I,J))\leqslant \exp(-\frac{1}{2M}k\log k\cdot\mu(I,J))$$
		for any $k\geqslant c'$.
		Given $c_3>0$, we choose $k'=\max\{c',t(I,J)\}$ where $t(I,J)\geqslant1$ is the unique solution of
		$$t\log t= c_3\frac{|J|}{\mu(I,J)}\log \frac{n}{|J|},$$
		then
		$$\begin{aligned}
		\mathbb{P}(e(I,J)\geqslant k'\mu(I,J))&\leqslant \exp(-\frac{1}{2M}k'\log k'\cdot\mu(I,J))\\
		&\leqslant\exp(-\frac{c_3}{2M}|J|\log \frac{n}{|J|}).
		\end{aligned}$$
		According to the proof of Lemma 4.2 in~\cite{lei2015}, we have
		\begin{equation}\label{appeq1}
			\mathbb{P}(\exists (I,J):|I|\leqslant|J|\leqslant n/\mathrm{e},e(I,J)\geqslant k'\mu(I,J))\leqslant n^{-(c_3/M-12)/2}.
		\end{equation}
		Therefore, when $c_3>12M$, for any $I,J\subset[n]$ with $|I|\leqslant|J|\leqslant n/\mathrm{e}$, with high probability at least one of the following holds:
		\begin{itemize}
			\item $k'=c'\geqslant t(I,J)$, which implies $e(I,J)\leqslant c'\mu(I,J)$.
			\item $k'=t(I,J)\geqslant c'$, which means $k'\log k'= c_3\frac{|J|}{\mu(I,J)}\log \frac{n}{|J|}$.
		\end{itemize}
		In the second case, according to~\eqref{appeq1}, we have $e(I,J)\leqslant k'\mu(I,J)$ with high probability, and then
		$$\frac{e(I,J)}{\mu(I,J)}\log \frac{e(I,J)}{\mu(I,J)}\leqslant k'\log k'=c_3\frac{|J|}{\mu(I,J)}\log \frac{n}{|J|},$$
		which yields
		$$e(I,J)\log \frac{e(I,J)}{\mu(I,J)}\leqslant c_3|J|\log \frac{n}{|J|}$$
		with high probability. Note that both $c_2$ and $c_3$ depend only on $c$ and $M$. The result follows by setting $c_2=\max\{c_1, c'\}$ and $c_3=(2c+12)M$.
												
		\begin{lemma}\label{thm2lm6}
			If both the bounded degree property and the bounded discrepancy property hold with some constant $c_1,c_2,c_3$, then
			$$\sup_{x,y\in T}|\sum_{(i,j)\in\bar{L}_{xy}}x_iy_jA_{ij}|=O(\sqrt{d}).$$
		\end{lemma}
		We refer to~\cite{lei2015,feige2005} for a proof.
												
		\section{Proof of Theorem~\ref{thm3}}\label{appC}
												
		Let $r>0$ be a fixed constant. We first provide several useful lemmas.
												
		\begin{lemma}\label{thm3lm1}
			The following two conclusions hold.
			\begin{itemize}
				\item[(i).]
				      There exists a constant $c_1=c_1(M,c_0,r)$ such that
				      $$\max_{l\in[n]}\lVert A^{(l)}-P\rVert\leqslant c_1\sqrt{d},\lVert A-P\rVert\leqslant c_1\sqrt{d}$$
				      with probability at least $1-O(n^{-r})$.
				      				      				      				      				      				
				\item[(ii).]
				      $\lVert H\rVert\leqslant1,\lVert H^{(l)}\rVert\leqslant1$. If $|\lambda_K|\geqslant 4\max\{\lVert A-P\rVert,\lVert A^{(l)}-P\rVert\}$, then $\lVert H-\mathrm{sgn}(H)\rVert^{1/2}\leqslant{2\lVert A-P\rVert}/{|\lambda_K|},\lVert H^{-1}\rVert\leqslant 2,\lVert (H^{(l)})^{-1}\rVert\leqslant 2$, and
				      \begin{equation}\label{appeq2}
				      	\lVert\hat{U}\hat{U}^T-\hat{U}^{(l)}(\hat{U}^{(l)})^T\rVert\leqslant\frac{2\lVert(A-A^{(l)})\hat{U}^{(l)}\rVert}{|\lambda_K|}.
				      \end{equation}
			\end{itemize}
		\end{lemma}
												
		\emph{Proof}:
		\begin{itemize}
			\item[(i).]
			      For any $l\in[n]$, it follows that
			      $$(A^{(l)}-P)_{ij}=\sum_{e\in E:a_{el}=0}\frac{a_{ei}(a_{ej}-\delta_{ij})}{|e|-1}(h_e-\mathbb{E}[h_e]).$$
			      According to the proof of Theorem~\ref{thm2}, there exists a constant $c_1>0$ such that
			      $\mathbb{P}(\lVert A-P\rVert\leqslant c_1\sqrt{d})\geqslant1-n^{-r}$ and
			      $\mathbb{P}(\lVert A^{(l)}-P\rVert\leqslant c_1\sqrt{d})\geqslant1-n^{-r-1}.$
			      Then
			      $$\begin{aligned}
			      &\mathbb{P}\left(\max_{l\in[n]}\lVert A^{(l)}-P\rVert\leqslant c_1\sqrt{d},\lVert A-P\rVert\leqslant c_1\sqrt{d}\right)\\
			      &\geqslant 1-n\cdot n^{-r-1}-n^{-r}=1-O(n^{-r}).
			\end{aligned}$$
															      			
			\item[(ii).]
			      We only prove~\eqref{appeq2}. See Lemma 2 in~\cite{abbe2020} and Lemma 4.14 in~\cite{chen2021} for the proof of the other conclusions.
			      			      			      			      			      			
			      According to Lemma 2.5 in~\cite{chen2021} and the Davis-Kahan theorem, we have
			      $$\lVert\hat{U}\hat{U}^T-\hat{U}^{(l)}(\hat{U}^{(l)})^T\rVert\leqslant\frac{\lVert(A-A^{(l)})\hat{U}^{(l)}\rVert}{\delta}$$
			      with $\delta=\min_{1\leqslant i\leqslant K<j\leqslant n}|\hat{\lambda}_i-\hat{\lambda}_j^{(l)}|$.
			      By Weyl's inequality, it follows that $|\hat{\lambda}_K-\lambda_K|\leqslant\lVert A-P\rVert\leqslant|\lambda_K|/4$. Then $|\hat{\lambda}_K|\geqslant 3/(4|\lambda_K|)$.
			      On the other hand, Weyl's inequality forces that
			      $$|\hat{\lambda}_{K+1}^{(l)}|=|\hat{\lambda}_{K+1}^{(l)}-\lambda_{K+1}|\leqslant\lVert A^{(l)}-P\lVert\leqslant|\lambda_K|/4.$$
			      Then $\delta=|\hat{\lambda}_{K}|-|\hat{\lambda}_{K+1}^{(l)}|\geqslant|\lambda_K|/2$, and the result follows.
			      			      			      			      			      			
		\end{itemize}
												
		\begin{lemma}\label{thm3lm2}
			For any fixed matrix $X\in\mathbb{R}^{n\times K}$, there exists a constant $C=C(M,c_0,r)$ such that
			$$\lVert (A-P)X\rVert_{2,\infty}\leqslant C\sqrt{d\log n}\lVert X\rVert_{2,\infty}$$
			with probability at least $1-O(n^{-r-1})$.
		\end{lemma}
												
		\emph{Proof}: We use the matrix Bernstein inequality to derive the bound. For each $i\in[n]$, we have
		$$\begin{aligned}
		\lVert(A-P)_{i\cdot}X\rVert&=\lVert\sum_{j=1}^n(A_{ij}-P_{ij})X_{j\cdot}\rVert\\
		&=\lVert\sum_{j=1}^n\left(\sum_{e\in E}\frac{a_{ei}(a_{ej}-\delta_{ij})}{|e|-1}(h_e-\mathbb{E}[h_e])\right)X_{j\cdot}\rVert\\
		&=\lVert\sum_{e\in E}\left(\sum_{j=1}^n\frac{a_{ei}(a_{ej}-\delta_{ij})}{|e|-1}X_{j\cdot}\right)(h_e-\mathbb{E}[h_e])\rVert
		.
		\end{aligned}$$
		Let $S_e=\left(\sum_{j=1}^n\frac{a_{ei}(a_{ej}-\delta_{ij})}{|e|-1}X_{j\cdot}\right)(h_e-\mathbb{E}[h_e])$.
		\begin{itemize}
			\item[(i).]
			      $\mathbb{E}[S_e]=0$, and
			      $$\begin{aligned}
			      \lVert S_e\rVert&\leqslant\lVert \sum_{j=1}^n\frac{a_{ei}(a_{ej}-\delta_{ij})}{|e|-1}X_{j\cdot}\rVert\\
			      &\leqslant \sum_{j=1}^n\frac{a_{ei}(a_{ej}-\delta_{ij})}{|e|-1}\lVert X_{j\cdot}\rVert\\
			      &\leqslant\sum_{j=1}^n\frac{a_{ei}(a_{ej}-\delta_{ij})}{|e|-1}\lVert X\rVert_{2,\infty}\\
			      &=a_{ei}\lVert X\rVert_{2,\infty}\leqslant M\lVert X\rVert_{2,\infty}.
			\end{aligned}$$
			\item[(ii).]
			      Define $W_e=\sum_{j=1}^n\frac{a_{ei}(a_{ej}-\delta_{ij})}{|e|-1}X_{j\cdot}$. Then
			      $$\begin{aligned}
			      v&=\max\{\lVert\sum_{e\in E}\mathbb{E}[S_eS_e^T]\rVert,\lVert\sum_{e\in E}\mathbb{E}[S_e^TS_e]\rVert\}\\
			      &=\max\{\lVert\sum_{e\in E}var(h_e)W_eW_e^T\rVert,\lVert\sum_{e\in E}var(h_e)W_e^TW_e\rVert\}\\
			      &\leqslant\max\{\sum_{e\in E}var(h_e)\lVert W_eW_e^T\rVert,\sum_{e\in E}var(h_e)\lVert W_e^TW_e\rVert\}\\
			      &=\sum_{e\in E}var(h_e)\lVert W_e\rVert^2.
			\end{aligned}$$
			Since $W_e$ is an $1\times K$ row vector, we have
			$$\begin{aligned}
			v&\leqslant\sum_{e\in E}var(h_e)\lVert\sum_{j=1}^n\frac{a_{ei}(a_{ej}-\delta_{ij})}{|e|-1}X_{j\cdot}\rVert^2\\
			&\leqslant\sum_{e\in E}\mathbb{E}[h_e]\left(\sum_{j=1}^n\frac{a_{ei}(a_{ej}-\delta_{ij})}{|e|-1}\lVert X_{j\cdot}\rVert\right)^2\\
			&\leqslant\sum_{e\in E}\mathbb{E}[h_e]\left(a_{ei}\lVert X\rVert_{2,\infty}\right)^2\\
			&=M\lVert X\rVert_{2,\infty}^2\sum_{e\in E}a_{ei}\mathbb{E}[h_e]\\
			&=M\lVert X\rVert_{2,\infty}^2\mathbb{E}[d_i]\leqslant Md\lVert X\rVert_{2,\infty}^2.
			\end{aligned}$$
		\end{itemize}
												
		Let $C>0$ be a constant such that $\frac{3C^2}{2M(3+C/\sqrt{c_0})}\geqslant r+3$. By the matrix Bernstein inequality, it follows that
		$$\begin{aligned}
		\mathbb{P}(\lVert(A-P)_{i\cdot}X\rVert&\geqslant C\sqrt{d\log n}\lVert X\rVert_{2,\infty})\\
		&\leqslant(1+K)\exp\left(\frac{-3C^2d\log n}{2M(3d+C\sqrt{d\log n})}\right)\\
		&\leqslant(1+K)\exp\left(\frac{-3C^2d\log n}{2M(3d+Cd/\sqrt{c_0})}\right)\\&=(1+K)n^{-\frac{3C^2}{2M(3+C/\sqrt{c_0})}}\\
		&\leqslant(1+K)n^{-r-3}\leqslant n^{-r-2}.
		\end{aligned}$$
		Then
		$$\mathbb{P}\left(\lVert(A-P)X\rVert_{2,\infty}\geqslant C\sqrt{d\log n}\lVert X\rVert_{2,\infty}\right)\leqslant n^{-r-1}.$$
												
		As a consequence, there exists a constant $c_2=c_2(M,c_0,r)$ such that
		$$\mathbb{P}(\lVert (A-P)U\rVert_{2,\infty}\leqslant c_2\sqrt{d\log n}\lVert U\rVert_{2,\infty})\geqslant1-O(n^{-r-1}).$$
		Furthermore, due to the independence between $(A-P)_{l\cdot}$ and $\hat{U}^{(l)}H^{(l)}-U$, by the proof of Lemma~\ref{thm3lm2}, there is a constant $c_3=c_3(M,c_0,r)$ such that for each $l\in[n]$
		$$\lVert(A-P)_{l\cdot}(\hat{U}^{(l)}H^{(l)}-U)\rVert\leqslant c_3\sqrt{d\log n}\lVert\hat{U}^{(l)}H^{(l)}-U\rVert_{2,\infty}$$
		with probability at least $1-O(n^{-r-1})$.
		Then
		$$\lVert(A-P)_{l\cdot}(\hat{U}^{(l)}H^{(l)}-U)\rVert\leqslant c_3\sqrt{d\log n}\lVert\hat{U}^{(l)}H^{(l)}-U\rVert_{2,\infty},$$
		for any $ l\in[n]$, with probability exceeding $1-O(n^{-r})$.
												
		\begin{lemma}\label{thm3lm3}
			For any fixed matrix $X\in\mathbb{R}^{n\times K}$, there exists a constant $C=C(M,c_0,r)$ such that
			$$\lVert (A-A^{(l)})X\rVert\leqslant C\sqrt{d\log n}\lVert X\rVert_{2,\infty},\forall l\in[n],$$
			with probability at least $1-O(n^{-r})$.
		\end{lemma}
												
		\emph{Proof}: We use the matrix Bernstein inequality to derive the bound. Define $W_e=\frac{a_ea_e^T-diag(a_e)}{|e|-1}$, then $A-A^{(l)}=\sum_{e\in E:a_{el}>0}(h_e-\mathbb{E}[h_e])W_e$ and
		$$\lVert(A-A^{(l)})X\rVert=\lVert\sum_{e\in E:a_{el}>0}(h_e-\mathbb{E}[h_e])W_eX\rVert.$$
		Define $S_e=(h_e-\mathbb{E}[h_e])W_eX$, then
		\begin{itemize}
			\item[(i).]
			      $\mathbb{E}[S_e]=0$, and
			      $$\begin{aligned}
			      \lVert S_e\rVert&\leqslant\lVert W_eX\rVert\\
			      &\leqslant\lVert W_eX\rVert_F\\
			      &=\sqrt{\sum_{i=1}^n\lVert\sum_{j=1}^n(W_e)_{ij}X_{j\cdot}\rVert^2}\\
			      &=\sqrt{\sum_{i=1}^n\lVert\sum_{j=1}^n\frac{a_{ei}(a_{ej}-\delta_{ij})}{|e|-1}X_{j\cdot}\rVert^2}\\
			      &\leqslant\sqrt{\sum_{i=1}^n\left(\sum_{j=1}^n\frac{a_{ei}(a_{ej}-\delta_{ij})}{|e|-1}\lVert X_{j\cdot}\rVert\right)^2}\\
			      &\leqslant \sqrt{\sum_{i=1}^n\left(a_{ei}\lVert X\rVert_{2,\infty}\right)^2}\\
			      &\leqslant\lVert X\rVert_{2,\infty}\sqrt{M\sum_{i=1}^na_{ei}}\\
			      &\leqslant M\lVert X\rVert_{2,\infty}.
			\end{aligned}$$
			\item[(ii).]
			      $$\begin{aligned}
			      v&=\max\{\lVert\sum_{e\in E:a_{el}>0}\mathbb{E}[S_eS_e^T]\rVert,\lVert\sum_{e\in E:a_{el}>0}\mathbb{E}[S_e^TS_e]\rVert\}\\
			      &\leqslant \sum_{e\in E:a_{el}>0}var(h_e)\lVert W_eX\rVert^2\\
			      &\leqslant\sum_{e\in E:a_{el}>0}\mathbb{E}[h_e]\lVert W_eX\rVert_F^2\\
			      &\leqslant\sum_{e\in E:a_{el}>0}\mathbb{E}[h_e]\cdot M^2\lVert X\rVert_{2,\infty}^2\\
			      &\leqslant M^2\lVert X\rVert_{2,\infty}^2\sum_{e\in E}a_{el}\mathbb{E}[h_e]\\&=M^2\lVert X\rVert_{2,\infty}^2\mathbb{E}[d_l]\\
			      &\leqslant M^2d\lVert X\rVert_{2,\infty}^2.
			\end{aligned}$$
		\end{itemize}
												
		Let $C>0$ be a constant such that $\frac{3C^2}{2M(3M+C/\sqrt{c_0})}\geqslant r+2$. By the matrix Bernstein inequality, it follows that
		$$\begin{aligned}
		\mathbb{P}(\lVert(A-A^{(l)})X\rVert&\geqslant C\sqrt{d\log n}\lVert X\rVert_{2,\infty})\\
		&\leqslant(1+K)\exp\left(\frac{-3C^2d\log n}{2M(3Md+C\sqrt{d\log n})}\right)\\
		&\leqslant(1+K)\exp\left(\frac{-3C^2d\log n}{2M(3Md+Cd/\sqrt{c_0})}\right)\\&=(1+K)n^{-\frac{3C^2}{2M(3M+C/\sqrt{c_0})}}\\
		&\leqslant(1+K)n^{-r-2}\leqslant n^{-r-1}.
		\end{aligned}$$
		Then
		$$\lVert (A-A^{(l)})X\rVert\leqslant C\sqrt{d\log n}\lVert X\rVert_{2,\infty},\forall l\in[n]$$
		with probability at least $1-n^{-r}$.

		Since $A-A^{(l)}$ is independent of $\hat{U}^{(l)}$, according to Lemma~\ref{thm3lm3}, there exists a constant $c_4=c_4(M,c_0,r)$ such that
		$$\lVert(A-A^{(l)})\hat{U}^{(l)}\rVert\leqslant c_4\sqrt{d\log n}\lVert \hat{U}^{(l)}\rVert_{2,\infty},\forall l\in[n]$$
		with probability exceeding $1-O(n^{-r})$.
												
		Define the following events:
		$$\begin{aligned}
		\mathcal{E}_1&=\{\lVert A^{(l)}-P\rVert\leqslant c_1\sqrt{d},\forall l\in[n],\text{and } \lVert A-P\rVert\leqslant c_1\sqrt{d}\},\\
		\mathcal{E}_2&=\{\lVert (A-P)U\rVert_{2,\infty}\leqslant c_2\sqrt{d\log n}\lVert U\rVert_{2,\infty}\},\\
		\mathcal{E}_3&=\{\lVert(A-P)_{l\cdot}(\hat{U}^{(l)}H^{(l)}-U)\rVert\\
		&\leqslant c_3\sqrt{d\log n}\lVert\hat{U}^{(l)}H^{(l)}-U\rVert_{2,\infty},\forall l\in[n]\},\\
		\mathcal{E}_4&=\{\lVert(A-A^{(l)})\hat{U}^{(l)}\rVert\leqslant c_4\sqrt{d\log n}\ \lVert \hat{U}^{(l)}\rVert_{2,\infty},\forall l\in[n]\}.
		\end{aligned}$$
												
		\begin{lemma}
			There exist some constants $c_5,c_6>0$ (depending only on $M,c_0$ and $r$) such that if $\sqrt{d\log n}\leqslant c_5|\lambda_K|$, then
			$$\begin{aligned}
			&\lVert A(\hat{U}H-U)\rVert_{2,\infty}\leqslant\\
			& c_6\left(\sqrt{d\log n}\lVert\hat{U}H-U\rVert_{2,\infty}+(\sqrt{d\log n}+\frac{\kappa d}{|\lambda_K|})\lVert U\rVert_{2,\infty}\right)
			\end{aligned}$$
			with probability at least $1-O(n^{-r})$.
		\end{lemma}
												
		\emph{Proof}: By the triangle inequality, it follows that
		$$\lVert A(\hat{U}H-U)\rVert_{2,\infty}\leqslant \lVert (A-P)(\hat{U}H-U)\rVert_{2,\infty}+\lVert P(\hat{U}H-U)\rVert_{2,\infty}.$$
												
		We first use the leave-one-out technique to develop an upper bound on $\lVert (A-P)(\hat{U}H-U)\rVert_{2,\infty}$. Applying the triangle inequality again yields
		$$\begin{aligned}
		&\lVert (A-P)(\hat{U}H-U)\rVert_{2,\infty}=\max_{l\in[n]}\ \lVert (A-P)_{l\cdot}(\hat{U}H-U)\rVert_{2}\\
		&\leqslant\max_{l\in[n]}\ (\lVert(A-P)_{l\cdot}(\hat{U}H-\hat{U}^{(l)}H^{(l)})\rVert\\
		&~~~+\lVert(A-P)_{l\cdot}(\hat{U}^{(l)}H^{(l)}-U)\rVert)\\
		&\leqslant\max_{l\in[n]}\ (\lVert A-P\rVert\lVert\hat{U}H-\hat{U}^{(l)}H^{(l)}\rVert\\
		&~~~+\lVert(A-P)_{l\cdot}(\hat{U}^{(l)}H^{(l)}-U)\rVert).
		\end{aligned}$$
		That is, $\hat{U}^{(l)}H^{(l)}$ is employed here as a surrogate of $\hat{U}H$.
												
		Set $c_5\leqslant1/(4c_1)$ such that $|\lambda_K|\geqslant 4c_1\sqrt{d}$. When $\mathcal{E}_1\bigcap\mathcal{E}_4$ happens, which has probability at least $1-O(n^{-r})$, we have
		$$\begin{aligned}
		\lVert \hat{U}H-\hat{U}^{(l)}H^{(l)}\rVert&=\lVert [\hat{U}\hat{U}^T-\hat{U}^{(l)}(\hat{U}^{(l)})^T]U\rVert\\
		&\leqslant\lVert \hat{U}\hat{U}^T-\hat{U}^{(l)}(\hat{U}^{(l)})^T\rVert\\&\leqslant\frac{2\lVert(A-A^{(l)})\hat{U}^{(l)}\rVert}{|\lambda_K|}\\&\leqslant \frac{2c_4\sqrt{d\log n}\lVert\hat{U}^{(l)}\rVert_{2,\infty}}{|\lambda_K|}\\
		&\leqslant {2c_4c_5\lVert\hat{U}^{(l)}\rVert_{2,\infty}}\\
		&\leqslant {4c_4c_5\lVert\hat{U}^{(l)}H^{(l)}\rVert_{2,\infty}}\\
		&\leqslant 4c_4c_5(\lVert\hat{U}H\rVert_{2,\infty}+\lVert\hat{U}H-\hat{U}^{(l)}H^{(l)}\rVert_{2,\infty})\\
		&\leqslant 4c_4c_5(\lVert\hat{U}H\rVert_{2,\infty}+\lVert\hat{U}H-\hat{U}^{(l)}H^{(l)}\rVert),
		\end{aligned}$$
		for all $l\in[n]$. So long as $4c_4c_5\leqslant\frac{1}{2}$, we have
		$$\begin{aligned}
		\lVert \hat{U}H-\hat{U}^{(l)}H^{(l)}\rVert&\leqslant 8c_4c_5\lVert\hat{U}H\rVert_{2,\infty}\\
		&\leqslant \lVert\hat{U}H\rVert_{2,\infty}\\
		&\leqslant\lVert\hat{U}H-U\rVert_{2,\infty}+\lVert U\rVert_{2,\infty},
		\end{aligned}$$
		for all $l\in[n]$. When $\mathcal{E}_1\bigcap\mathcal{E}_3\bigcap\mathcal{E}_4$ happens, which has probability at least $1-O(n^{-r})$, we can deduce that for all $l\in[n]$,
		$$\begin{aligned}
		&\lVert(A-P)_{l\cdot}(\hat{U}^{(l)}H^{(l)}-U)\rVert\leqslant c_3\sqrt{d\log n}\lVert\hat{U}^{(l)}H^{(l)}-U\rVert_{2,\infty}\\
		&\leqslant c_3\sqrt{d\log n}\left(\lVert\hat{U}H-\hat{U}^{(l)}H^{(l)}\rVert_{2,\infty}+\lVert\hat{U}H-U\rVert_{2,\infty}\right)\\
		&\leqslant c_3\sqrt{d\log n}\left(\lVert\hat{U}H-\hat{U}^{(l)}H^{(l)}\rVert+\lVert\hat{U}H-U\rVert_{2,\infty}\right)\\
		&\leqslant c_3\sqrt{d\log n}\left(2\lVert\hat{U}H-U\rVert_{2,\infty}+\lVert U\rVert_{2,\infty}\right).
		\end{aligned}$$
		Combining these two results, we have
		$$\begin{aligned}
		&\lVert (A-P)(\hat{U}H-U)\rVert_{2,\infty}\\
		&\leqslant\max_{l\in[n]}\ (\lVert A-P\rVert\lVert\hat{U}H-\hat{U}^{(l)}H^{(l)}\rVert
		+\lVert(A-P)_{l\cdot}(\hat{U}^{(l)}H^{(l)}-U)\rVert)\\
		&\leqslant c_1\sqrt{d}(\lVert\hat{U}H-U\rVert_{2,\infty}+\lVert U\rVert_{2,\infty})\\
		&~~~+c_3\sqrt{d\log n}\left(2\lVert\hat{U}H-U\rVert_{2,\infty}+\lVert U\rVert_{2,\infty}\right)\\
		&\leqslant(c_1+c_3)\sqrt{d\log n}\left(3\lVert\hat{U}H-U\rVert_{2,\infty}+2\lVert U\rVert_{2,\infty}\right)
		\end{aligned}$$
		with probability exceeding $1-O(n^{-r})$.
												
		Next, we bound $\lVert P(\hat{U}H-U)\rVert_{2,\infty}$. According to Eq.~(4.114) in~\cite{chen2021}, when $\mathcal{E}_1$ happens and $|\lambda_K|\geqslant4c_1\sqrt{d}$, we have
		$$\begin{aligned}
		\lVert P(\hat{U}H-U)\rVert_{2,\infty}&\leqslant \lVert U\rVert_{2,\infty}\cdot\lVert \Lambda\rVert\cdot\lVert\hat{U}\hat{U}^T-UU^T\rVert^2\\
		&\leqslant\lVert U\rVert_{2,\infty}\cdot\lVert \Lambda\rVert\cdot\frac{ 2\lVert A-P\rVert^2}{\lambda_K^2}\\
		&\leqslant|\lambda_1|\cdot\frac{2c_1^2d}{\lambda_K^2}\lVert U\rVert_{2,\infty}\\
		&=\frac{2c_1^2\kappa d}{|\lambda_K|}\lVert U\rVert_{2,\infty}.
		\end{aligned}$$
												
		Combining these two bounds, the result follows by letting $c_5=\min\{1/(4c_1),1/(8c_4)\}$ and $c_6=\max\{3(c_1+c_3),2(c_1+c_3)+2c_1^2\}$.
												
		Define
		$
		\mathcal{E}_5=\{\lVert A(\hat{U}H-U)\rVert_{2,\infty}\leqslant c_6(\sqrt{d\log n}\lVert\hat{U}H-U\rVert_{2,\infty}+(\sqrt{d\log n}+\kappa d/|\lambda_K|)\lVert U\rVert_{2,\infty})\}.
		$
		We are now ready to prove Theorem~\ref{thm3}.
												
		\textbf{\emph{Proof}}:
		Suppose $\sqrt{d\log n}\leqslant c_5|\lambda_K|$ and $\bigcap_{i=1}^5\mathcal{E}_i$ happens, which has probability exceeding $1-O(n^{-r})$. According to Lemma 4.16 in~\cite{chen2021}, we have
		$$\lVert\hat{U}H-U\rVert_{2,\infty}\leqslant\gamma_1+\gamma_2+\gamma_3$$
		with $\gamma_1=\frac{2\lVert A(\hat{U}H-U)\rVert_{2,\infty}}{|\lambda_K|}$, $\gamma_2=\frac{4c_1\sqrt{d}\lVert AU\rVert_{2,\infty}}{\lambda_K^2}$, and $\gamma_3=\frac{\lVert(A-P)U\rVert_{2,\infty}}{|\lambda_K|}$. The last two terms could be directly bounded.
												
		First, when $\mathcal{E}_2$ happens, we have
		$$\gamma_3\leqslant c_2\frac{\sqrt{d\log n}}{|\lambda_K|}\lVert U\rVert_{2,\infty}.$$
		Next, since
		$$\begin{aligned}
		\lVert AU\rVert_{2,\infty}&\leqslant\lVert (A-P)U\rVert_{2,\infty}+\lVert PU\rVert_{2,\infty}\\
		&= \lVert (A-P)U\rVert_{2,\infty}+\lVert U\Lambda\rVert_{2,\infty}\\
		&\leqslant c_2\sqrt{d\log n}\lVert U\rVert_{2,\infty}+\lVert U\rVert_{2,\infty}\lVert\Lambda\rVert\\
		&= (c_2\sqrt{d\log n}+|\lambda_1|)\lVert U\rVert_{2,\infty},
		\end{aligned}$$
		we have
		$$\begin{aligned}
		\gamma_2&\leqslant 4c_1(c_2\frac{\sqrt{d\log n}}{|\lambda_K|}+\kappa)\frac{\sqrt{d}}{|\lambda_K|}\lVert U\rVert_{2,\infty}\\
		&\leqslant 4c_1(c_2c_5+\kappa)\frac{\sqrt{d}}{|\lambda_K|}\lVert U\rVert_{2,\infty}.
		\end{aligned}$$
		Now we turn to bound $\gamma_1$. For large $n$ such that $c_5\kappa\leqslant\log n$, we have
		$$\begin{aligned}
		&\gamma_1=\frac{2}{|\lambda_K|}\lVert A(\hat{U}H-U)\rVert_{2,\infty}\\
		&\leqslant{2c_6}\frac{\sqrt{d\log n}}{|\lambda_K|}\left(\lVert\hat{U}H-U\rVert_{2,\infty}+(1+\frac{\kappa \sqrt{d}}{\sqrt{\log n}|\lambda_K|})\lVert U\rVert_{2,\infty}\right)\\
		&\leqslant{2c_6}\frac{\sqrt{d\log n}}{|\lambda_K|}\left(\lVert\hat{U}H-U\rVert_{2,\infty}+\left(1+\frac{c_5\kappa}{\log n}\right)\lVert U\rVert_{2,\infty}\right)\\
		&\leqslant{2c_6}\frac{\sqrt{d\log n}}{|\lambda_K|}\left(\lVert\hat{U}H-U\rVert_{2,\infty}+2\lVert U\rVert_{2,\infty}\right).
		\end{aligned}$$
		The upper bound on $\lVert\hat{U}H-U\rVert_{2,\infty}$ gives that
		$$\begin{aligned}
		\lVert\hat{U}H-U\rVert_{2,\infty}&\leqslant {2c_6}\frac{\sqrt{d\log n}}{|\lambda_K|}\left(\lVert\hat{U}H-U\rVert_{2,\infty}+2\lVert U\rVert_{2,\infty}\right)\\
		&~~~+\gamma_2+\gamma_3.
		\end{aligned}$$
		As long as $c_5\cdot2c_6\leqslant1/2$, we have
		$$\begin{aligned}
		&\lVert\hat{U}H-U\rVert_{2,\infty}\leqslant {8c_6}\frac{\sqrt{d\log n}}{|\lambda_K|}\lVert U\rVert_{2,\infty}+2\gamma_2+2\gamma_3\\
		&\leqslant\left({8c_6}+\frac{8c_1(c_2c_5+\kappa)}{\sqrt{\log n}}+2c_2\right)\frac{\sqrt{d\log n}}{|\lambda_K|}\lVert U\rVert_{2,\infty}.
		\end{aligned}$$
		Recall that $\kappa=O(\log n)$ by Assumption~\ref{asp3}. Let the constant $c_7=c_7(M,c_0,r)$ be an upper bound on ${8c_6}+\frac{8c_1(c_2c_5+\kappa)}{\sqrt{\log n}}+2c_2$. When $\log n>1$, we have
		$$\begin{aligned}
		&\lVert\hat{U}\mathrm{sgn}(H)-\hat{U}H\rVert_{2,\infty}\leqslant\lVert\hat{U}\rVert_{2,\infty}\lVert H-\mathrm{sgn}(H)\rVert\\&\leqslant2\lVert\hat{U}H\rVert_{2,\infty}\lVert H-\mathrm{sgn}(H)\rVert\\&\leqslant2(\lVert\hat{U}H-U\rVert_{2,\infty}+\lVert U\rVert_{2,\infty})\cdot\frac{4\lVert A-P\rVert^2}{\lambda_K^2}\\&\leqslant\frac{8c_1^2d}{\lambda_K^2}(\lVert\hat{U}H-U\rVert_{2,\infty}+\lVert U\rVert_{2,\infty})\\&\leqslant8c_1^2c_5(1+c_5c_7)\frac{\sqrt{d\log n}}{|\lambda_K|}\lVert U\rVert_{2,\infty}.
		\end{aligned}$$
		Applying the triangle inequality yields
		$$\lVert\hat{U}\mathrm{sgn}(H)-U\rVert_{2,\infty}\leqslant\lVert\hat{U}\mathrm{sgn}(H)-\hat{U}H\rVert_{2,\infty}+\lVert\hat{U}H-U\rVert_{2,\infty}.$$
		Thus, the result follows by choosing $C_3=\min\{c_5,1/(4c_6)\}$ and $C_4=c_7+8c_1^2c_5(1+c_5c_7)$.
		$\hfill\blacksquare$

		\section*{Acknowledgment}
		This work was supported by the Natural Science Foundation of China (No. 12071281) and the National Key R \& D Program of China (No. 2021YFA1003004).

		\ifCLASSOPTIONcaptionsoff
		\newpage
		\fi

												
												

\begin{thebibliography}{27}
																		
			\bibitem{holland1983}
			P. W. Holland, K. B. Laskey, and S. Leinhardt, \lq\lq Stochastic blockmodels: first steps,\rq\rq~\textit{Social Netw.}, vol. 5, no. 2, pp. 109--137, 1983.
																		
			\bibitem{karrer2011}
			B. Karrer, and M. E. J. Newman, \lq\lq Stochastic blockmodels and community structure in networks,\rq\rq~\textit{Phys. Rev. E}, vol. 83, no. 1, 2011, Art. no. 016107.
																		
			\bibitem{zhao2012}
			Y. Zhao, E. Levina, and J. Zhu, \lq\lq Consistency of community detection in networks under degree-corrected stochastic block models,\rq\rq~\textit{Ann. Statist.}, vol. 40, no. 4, pp. 2266--2292, 2012.
																		
			\bibitem{krzakala2013}
			F. Krzakala \textit{et al.}, \lq\lq Spectral redemption in clustering sparse networks,\rq\rq~\textit{Proc. Nat. Acad. Sci. USA}, vol. 110, no. 52, pp. 20935--20940, 2013.
																		
			\bibitem{gao2018}
			C. Gao, Z. Ma, A. Y. Zhang, and H. H. Zhou, \lq\lq Community detection in degree-corrected block models,\rq\rq~\textit{Ann. Statist.}, vol. 46, no. 5, pp. 2153--2185, 2018.
																		
			\bibitem{gulikers2018}
			L. Gulikers, M. Lelarge, and L. Massouli\'{e}, \lq\lq An impossibility result for reconstruction in the degree-corrected stochastic block model,\rq\rq~\textit{Ann. Appl. Probab.}, vol. 28, no. 5, pp. 3002--3027, 2018.
																		
			\bibitem{rohe2011}
			K. Rohe, S. Chatterjee, and B. Yu, \lq\lq Spectral clustering and the high-dimensional stochastic block model,\rq\rq~\textit{Ann. Statist.}, vol. 39, no. 4, pp. 1878--1915, 2011.
																		
			\bibitem{lei2015}
			J. Lei, and A. Rinaldo, \lq\lq Consistency of spectral clustering in stochastic block models,\rq\rq~\textit{Ann. Statist.}, vol. 43, no. 1, pp. 215--237, 2015.
																		
			\bibitem{chodrow2021}
			P. S. Chodrow, N. Veldt, and A. R. Benson, \lq\lq Generative hypergraph clustering: from blockmodels to modularity,\rq\rq~\textit{Sci. Adv.}, vol. 7, 2021, Art. no. eabh1303.
																		
			\bibitem{ahn2018}
			K. Ahn, K. Lee, and C. Suh, \lq\lq Hypergraph spectral clustering in the weighted stochastic block model,\rq\rq~\textit{IEEE J. Sel. Topics Signal Process.}, vol. 12, no. 5, pp. 959--974, 2018.
																		
																		
			\bibitem{ghoshdastidar2017}
			D. Ghoshdastidar, and A. Dukkipati, \lq\lq Consistency of spectral hypergraph partitioning under planted partition model,\rq\rq~\textit{Ann. Statist.}, vol. 45, no. 1, pp. 289--315, 2017.
																		
			\bibitem{ke2019}
			Z. T. Ke, F. Shi, and D. Xia, \lq\lq Community detection for hypergraph networks via regularized tensor power iteration,\rq\rq~arXiv:1909.06503, 2019.
																		
			\bibitem{cole2020}
			S. Cole, and Y. Zhu, \lq\lq Exact recovery in the hypergraph stochastic block model: a spectral algorithm,\rq\rq~\textit{Linear Algebra Appl.}, vol. 593, pp. 45--73, 2020.
																		
			\bibitem{chien2019}
			I. E. Chien, C.-Y. Lin, and I.-H. Wang, \lq\lq On the minimax misclassification ratio of hypergraph community detection,\rq\rq~\textit{IEEE Trans. Inf. Theory}, vol. 65, no. 12, pp. 8095--8118, 2019.
																		
			\bibitem{zhang2021}
			Q. Zhang, and V. Y. Tan, \lq\lq Exact recovery in the general hypergraph stochastic block model,\rq\rq~\textit{IEEE Trans. Inf. Theory}, vol. 69, no. 1, pp. 453--471, 2022.
																		
			\bibitem{su2019}
			L. Su, W. Wang, and Y. Zhang, \lq\lq Strong consistency of spectral clustering for stochastic block models,\rq\rq~\textit{IEEE Trans. Inf. Theory}, vol. 66, no. 1, pp. 324--338, 2019.
																		
			\bibitem{abbe2020}
			E. Abbe, J. Fan, K. Wang, and Y. Zhong, \lq\lq Entrywise eigenvector analysis of random matrices with low expected rank,\rq\rq~\textit{Ann. Statist.}, vol. 48, no. 3, pp. 1452--1474, 2020.
																		
			\bibitem{kim2018}
			C. Kim, A. S. Bandeira, and M. X. Goemans, \lq\lq Stochastic block model for hypergraphs: statistical limits and a semidefinite programming approach,\rq\rq~arXiv:1807.02884, 2018.
																		
			\bibitem{Gaudio2022}
			J. Gaudio, and N. Joshi, \lq\lq Community detection in the hypergraph SBM: optimal recovery given the similarity matrix,\rq\rq~arXiv:2208.12227, 2022.
											
			\bibitem{Wang2023}
			H. Wang, \lq\lq Strong consistency and optimality of spectral clustering in symmetric binary non-uniform hypergraph stochastic block model,\rq\rq~arXiv:2306.06845, 2023.
			
			\bibitem{Dumitriu2023}
			I. Dumitriu and H. Wang, \lq\lq Exact recovery for the non-uniform hypergraph stochastic block model,\rq\rq~arXiv:2304.13139, 2023.
			
			\bibitem{carletti2021}
			T. Carletti, D. Fanelli, and R. Lambiotte, \lq\lq Random walks and community detection in hypergraphs,\rq\rq~\textit{J. Phys. Complex}, vol. 2, no. 1, 2021, Art. no. 015011.
																		
			\bibitem{abbe2017}
			E. Abbe, \lq\lq Community detection and stochastic block models: recent developments,\rq\rq~\textit{J. Mach. Learn. Res.}, vol. 18, no. 1, pp. 6446--6531, 2017.
										
			\bibitem{Friedman1989}
			J. Friedman, J. Kahn, and E. Szemeredi, \lq\lq On the second eigenvalue of random regular graphs,\rq\rq~in \textit{Proc. 21st Annu. ACM Symp. Theory Comput.}, 1989, pp. 587--598.
			
			
			\bibitem{feige2005}
			U. Feige, and E. Ofek, \lq\lq Spectral techniques applied to sparse random graphs,\rq\rq~\textit{Random Struct. Algorithms}, vol. 27, no. 2, pp. 251--275, 2005.
																		
			\bibitem{kumar2004simple}
			A. Kumar, Y. Sabharwal, and S. Sen, \lq\lq A simple linear time (1+$\epsilon$)-approximation algorithm for $k$-means clustering in any dimensions,\rq\rq~\textit{45th Annual IEEE Symp. Found. Comput. Sci.}, 2004, pp. 454--462.
																		
			\bibitem{gross2011}
			D. Gross, \lq\lq Recovering low-rank matrices from few coefficients in any basis,\rq\rq~\textit{IEEE Trans. Inf. Theory}, vol. 57, no. 3, pp. 1548--1566, 2011.
																		
			\bibitem{tropp2015introduction}
			J. A. Tropp, \lq\lq An introduction to matrix concentration inequalities,\rq\rq~\textit{Found. Trends Mach. Learn.}, vol. 8, no. 1, pp. 1--230, 2015.
																		
			\bibitem{mao2021}
			X. Mao, P. Sarkar, and D. Chakrabati, \lq\lq Estimating mixed memberships with sharp eigenvector deviations,\rq\rq~\textit{J. Amer. Stat. Assoc.}, vol. 116, no. 536, pp. 1928--1940, 2021.
																		
			\bibitem{chen2021}
			Y. Chen, Y. Chi, J. Ffan, and C. Ma, \lq\lq Spectral methods for data science: a statistical perspective,\rq\rq~\textit{Found. Trends Mach. Learn.}, vol. 14, no. 5, pp. 566--806, 2021.
																		
		\end{thebibliography}
		%

		%
												
		\vspace{-300pt}
												
		\begin{IEEEbiographynophoto}{Chong Deng}
			received the B.Sc. degree in applied mathematics from Shanghai University in 2020. He is currently working toward the M.S. degree in applied mathematics at Shanghai University. His research interests are the theory and applications of stochastic block models.
												
		\end{IEEEbiographynophoto}
				
		\vspace{-320pt}
												
												
		\begin{IEEEbiographynophoto}{Xin-Jian Xu}
			received the B.Sc. and Ph.D. degerees in theoretical physics from Lanzhou University in 2001 and 2005 respectively. He is currently a professor in Qianweichang College at Shanghai University. His research interests include statistical analysis of complex networks and its applications in society and biology.
		\end{IEEEbiographynophoto}
						
		\vspace{-320pt}
						
		\begin{IEEEbiographynophoto}{Shihui Ying}
			(Member, IEEE) received the B.Eng. degree in mechanical engineering and the Ph.D. degree in applied mathematics from Xi'an Jiaotong University in 2001 and 2008 respectively. He is currently a professor in the Department of Mathematics at Shanghai University. He was a postdoctor in Biomedical Research Imaging Center (BRIC) at University of North Carolina at Chapel Hill from 2012 to 2013. His research interests cover geometric theory and methods for machine learning with applications in medical image analysis.
		\end{IEEEbiographynophoto}				
												

\end{document}